# Graph and Network Theory

Ernesto Estrada
Department of Mathematics and Statistics
University of Strathclyde, Glasgow

## Introduction

Graph Theory was born in 1736 when Leonhard Euler published "*Solutio problematic as geometriam situs pertinentis*" (The solution of a problem relating to the theory of position) (Euler, 1736). This history is well documented (Biggs et al., 1976) and widely available in any textbook of graph or network theory. However, the word graph appeared for the first time in the context of natural sciences in 1878, when the English mathematician James J. Sylvester wrote a paper entitled "Chemistry and Algebra" which was published in *Nature* (Sylvester, 1877-78), where he wrote that "*Every invariant and covariant thus becomes expressible by a **graph** precisely identical with a Kekulean diagram or chemicograph*". The use of graph theory in condensed matter physics, pioneered by many chemical and physical graph theorists (Harary, 1968; Trinajstić, 1992), is today well established; it has become even more popular after the recent discovery of graphene.

There are few, if any, areas of physics in the XXIst century in which graphs and network are not involved directly or indirectly. Hence it is impossible to cover all of them in this Chapter. Thus I owe the reader an apology for the incompleteness of this Chapter and a promise to write a more complete treatise. For instance, quantum graphs are not considered in this Chapter and the reader is referred to a recent introductory monograph on this topic for details (Berkolaiko, Kuchment, 2013). In this chapter we will cover some of the most important areas of applications of graph theory in physics. These include condensed matter physics, statistical physics, quantum electrodynamics, electrical networks and vibrational problems. In the second part we summarise some of the most important aspects of the study of complex networks. This is an interdisciplinary area which has emerged with tremendous impetus in the XXIst century which studies networks appearing in complex systems. These systems range from molecular and biological to ecological, social and technological systems. Thus graph theory and network theory have helped to broaden the horizons of physics to embrace the study of new complex systems.

We hope this chapter motivates the reader to find more about the connections between graph/network theory and physics, consolidating this discipline as an important part of the curriculum for the physicists of the XXIst century.

# 1 The language of graphs and networks

The first thing that needs to be clarified is that the terms graphs and networks are used indistingtly in the literature. In this Chapter we will reserve the term graph for the abstract mathematical concept, in general referred to small, artificial formations of nodes and edges. The term network is then reserved for the graphs representing real-world objects in which the nodes represent entities of the system and the edges represent the relationships among them. Therefore, it is clear that we will refer to the system of individuals and their interactions as a 'social network' and not as a 'social graph'. However, they should mean exactly the same.

For the basic concepts of graph theory the reader is recommended to consult the introductory book by Harary (1967). We start by defining a graph formally. Let us consider a finite set $V = \{v_1, v_2, \cdots, v_n\}$ of unspecified elements and let $V \otimes V$ be the set of all ordered pairs $[v_i, v_j]$ of the elements of $V$. A relation on the set $V$ is any subset $E \subseteq V \otimes V$. The relation $E$ is symmetric if $[v_i, v_j] \in E$ implies $[v_j, v_i] \in E$ and it is reflexive if $\forall v \in V, [v, v] \in E$. The relation $E$ is antireflexive if $[v_i, v_j] \in E$ implies $v_i \neq v_j$. Now we can define a **simple graph** as the pair $G = (V, E)$, where $V$ is a finite set of nodes, vertices or points and $E$ is a symmetric and antireflexive relation on $V$, whose elements are known as the edges or links of the graph. In a **directed graph** the relation $E$ is non-symmetric. In many physical applications the edges of the graphs are required to support weights, i.e., real numbers indicating a specific property of the edge. In this case the following more general definition is convenient. A **weighted graph** is the quadruple $G = (V, E, W, f)$ where $V$ is a finite set of nodes, $E \subseteq V \otimes V = \{e_1, e_2, \cdots, e_m\}$ is a set of edges, $W = \{w_1, w_2, \cdots, w_r\}$ is a set of weights such that $w_i \in \mathbb{R}$ and $f: E \to W$ is a surjective mapping that assigns a weight to each edge. If the weights are natural numbers then the resulting graph is a multigraph in which there could be multiple edges between pairs of vertices. That is, if the weight between nodes $p$ and $q$ is $k \in N$ it means that there are $k$ links between the two nodes.

In an undirected graph we say that wo nodes $p$ and $q$ are adjacent if they are joined by an edge $e = \{p, q\}$. In this case we say that the nodes $p$ and $q$ are incident to the link $e$, and the link $e$ is incident to the nodes $p$ and $q$. The two nodes are called the end nodes of the edge. Two edges $e_1 = \{p, q\}$ and $e_2 = \{r, s\}$ are adjacent if they are both incident to at least one node. A simple but important characteristic of a node is its degree, which is defined as the number of edges which are incident to it or similarly the number of nodes adjacent to it. Slightly different definitions apply to directed graphs. The node $p$ is adjacent to node $q$ if there is a directed link from $p$ to $q$, $e = (p, q)$. We also say that a link from $p$ to $q$ is incident from $p$ and incident to $q$; $p$ is incident to $e$ and $q$ is incident from $e$. Consequently, we have two different kinds of degrees in directed graphs. The in-degree of a node is the number of links incident to it and its out-degree is the number of links incident from it.

## 1.1 Graph operators

The incidence and adjacency relations in graphs allow us to define the following graph operators. We consider an undirected graph for which we construct its **incidence matrix** with an arbitrary orientation of its entries. This is necessary to consider that the incidence matrix is a discrete analogous of the gradient. That is, for every edge $\{p,q\}$, $p$ is the positive (head) and $q$ the negative (tail) end of the oriented link. Let the links of the graph be labeled as $e_1, e_2, \cdots, e_m$. Hence the **oriented incidence matrix** $\nabla(G)$:

$$\nabla_{ij}(G) = \begin{cases} +1 & \text{node } v_i \text{ is the head of link } e_j \\ -1 & \text{node } v_i \text{ is the tail of link } e_j \\ 0 & \text{otherwise} \end{cases}$$

We remark that the results obtained below are independent of the orientation of the links but assume that once the links are oriented, this orientation is not changed. Let the vertex $L_V$ and edge $L_E$ spaces be the vector spaces of all real-valued functions defined on $V$ and $E$, respectively. The **incidence operator** of the graph is then defined as

$$\nabla(G): L_V \to L_E, \qquad (1.1)$$

such that for an arbitrary function $f: V \to \mathbb{R}$, $\nabla(G)f: E \to \Re$ is given by

$$(\nabla(G)f)(e) = f(p) - f(q), \qquad (1.2)$$

where $p$ are the starting (head) and $q$ the ending (tail) points of the oriented link $e$. Here we consider that $f$ is a real or vector-valued function on the graph with $|f|$ being $\mu$-measurable for certain measure $\mu$ on the graph.

On the other hand, let $H$ be a Hilbert space with scalar product $\langle \cdot, \cdot \rangle$ and norm $\|\cdot\|$. Let $G = (V, E)$ be a simple graph. The **adjacency operator** is an operator acting on the Hilbert space $H := l^2(V)$ defined as

$$(\mathbf{A}f)(p) := \sum_{u,v \in E} f(q), \quad f \in H, \quad i \in V. \qquad (1.3)$$

The adjacency operator of an undirected network is a *self-adjoint operator*, which is bounded on $l^2(V)$. We recall that $l^2$ is the Hilbert space of square summable sequences with inner product, and that an operator is self-adjoint if its matrix is equal to its own conjugate transpose, i.e., it is Hermitian. It is worth pointing out here that the adjacency operator of a directed network might not be self-adjoint. The matrix representation of this operator is the adjacency matrix $\mathbf{A}$, which for a simple graph is defined as

$$A_{ij} = \begin{cases} 1 & \text{if } i, j \in E \\ 0 & \text{otherwise.} \end{cases} \qquad (1.4)$$

A third operator which is related to the previous two and which plays a fundamental role in the applications of graph theory in physics is the **Laplacian operator**. This operator is defined by

$$\mathbf{L}(G)f = -\nabla \cdot (\nabla f), \qquad (1.5)$$

and it is the graph version of the Laplacian operator

$$\Delta f = \frac{\partial^2 f}{\partial x_1^2} + \frac{\partial^2 f}{\partial x_2^2} + \cdots + \frac{\partial^2 f}{\partial x_n^2}. \qquad (1.6)$$

The negative sign in (1.5) is used by convention. Then the Laplacian operator acting on the function $f$ previously defined is given by

$$(\mathbf{L}(G)f)(u) = \sum_{\{u,v\} \in E} [f(u) - f(v)], \qquad (1.7)$$

which in matrix form is given by

$$L_{uv}(G) = \sum_{e \in E} \nabla_{eu} \nabla_{ev} = \begin{cases} -1 & \text{if } uv \in E, \\ k_u & \text{if } u = v, \\ 0 & \text{otherwise.} \end{cases} \qquad (1.8)$$

Using the degree matrix **K** which is a diagonal matrix of the degrees of the nodes in the graph, the Laplacian and adjacency matrices of a graph are related by

$$\mathbf{L} = \mathbf{K} - \mathbf{A}. \qquad (1.9)$$

## 1.2 General graph concepts

Other important general concepts of graphs theory which are fundamental for the study of graphs and networks in physics are the following. Two graphs $G_1$ and $G_2$ are **isomorphic** if there is a one-to-one correspondence between the nodes of $G_1$ and those of $G_2$, such as the number of edges joining each pair of nodes in $G_1$ is equal to that joining the corresponding pair of nodes in $G_2$. If the graphs are directed the edges must coincide not only in number but also in direction. The graph $S = (V', E')$ is a **subgraph** of a graph $G = (V, E)$ if and only if $V' \subseteq V$ and $E' \subseteq E$. A particular kind of subgraph is the **clique**, which is a maximal complete subgraph of a graph. A **complete graph** is the one in which every pair of nodes are connected. A (directed) **walk** of length $L$ from $v_1$ to $v_{L+1}$ is any sequence of (not necessarily different) nodes $v_1, v_2, \cdots, v_L, v_{L+1}$ such that for each $i = 1, 2, \cdots, L$ there is link from $v_i$ to $v_{i+1}$. A walk is closed (CW) if $v_{L+1} = v_1$. A particular kind of walk is the **path** of length $L$, which is a walk of length $L$ in which all the nodes (and all the adges) are distinct. A **trial** has all the links different but not necessarily all the nodes. A **cycle** is a closed walk in which all the edges and all the nodes (except the first and last) are distinct. The **girth** of the graph is the size (number of nodes) of the minimum cycle in the graph.

A graph is **connected** if there is a path between any pair of nodes in the graph. Otherwise it is disconnected. Every connected subgraph is a **connected component** of the graph. The analogous concept in a directed graph is that of **strongly connected** graph. A directed graph is strongly connected if there is a directed path between each pair of nodes. The strongly connected components of a directed graph are its maximal strongly connected subgraphs.

In an undirected graph the **shortest path distance** $d(p, q) = d_{pq}$ is the number of edges in the shortest path between the nodes $p$ and $q$ in the graph. If $p$ and $q$ are in different connected components of the graph the distance between them is set to infinite, $d(p, q) := \infty$. In a directed graph it is typical to consider the *directed distance* $\vec{d}(p, q)$ between a pair of nodes $p$ and $q$ as the length of the directed shortest path from $p$ to $q$. However, in general $\vec{d}(p, q) \neq \vec{d}(q, p)$, which violates the symmetry property of a metric, so that $\vec{d}(p, q)$ is not a distance but a *pseudo-distance* or a *pseudo-metric*. The distance between all pairs of nodes in a graph can be arranged in a distance matrix **D** which for undirected graphs is a square symmetric matrix. The maximum entry for a given row/column of the distance matrix of an undirected (strongly connected directed) graph is known as the **eccentricity** $e(p)$ of the node $p$, $e(p) = \max_{x \in V(G)} \{d(p, x)\}$. The maximum eccentricity among the nodes of a graph is the **diameter** of the graph, which is $diam(G) = \max_{x,y \in V(G)} \{d(x,y)\}$. The **average path length** $\bar{l}$ of a graph with $n$ node is

$$\bar{l} = \frac{1}{n(n-1)} \sum_{x,y} d(x,y).  \qquad (1.10)$$

An important measure for the study of networks was introduced by Watts and Strogatz (1998) as a way of quantifying how clustered a node is. For a given node $i$ the **clustering coefficient** is the number of triangles connected to this node $|C_3(i)|$ divided by the number of triples centred on it

$$C_i = \frac{2|C_3(i)|}{k_i(k_i - 1)}, \qquad (1.11)$$

where $k_i$ is the degree of the node. The average value of the clustering for all nodes in a network $\bar{C}$

$$\bar{C} = \frac{1}{n} \sum_{i=1}^{n} C_i \qquad (1.12)$$

has been extensively used in the analysis of complex networks (see Section 8 of this Chapter).

A second clustering coefficient has been introduced as a global characterization of network cliquishness (Newman et al., 2001). This index which is also known as **network transitivity**, is defined as the ratio of three times the number of triangles divided by the number of connected triples (2-paths):

$$C = \frac{3|C_3|}{|P_2|}. \qquad (1.13)$$

## 1.3 Types of graphs

The simplest type of graph is the tree. A **tree** of $n$ nodes is a graph which is connected and has no cycles. The simplest tree is the **path** $P_n$. The path (also know as linear path or chain) is the tree of $n$ nodes, $n-2$ of which have degree 2 and two nodes have degree 1. For any kind of graph we can find a spanning tree, which is a subgraph of this graph that includes every node and is a tree. A **forest** is a disconnected graph in which every connected component is a tree. A **spanning forest** is a subgraph of the graph that includes every node and is a forest.

An $r$-**regular graph** is a graph with $rn/2$ edges in which all nodes have degree $r$. A particular case of regular graph is the complete graph previously defined. Another type of regular graph is the **cycle**, which is a regular graph of degree 2, i.e., a $2$-regular graph, denoted by $C_n$. The **complement** of a graph $G$ is the graph $\bar{G}$ with the same set of nodes as $G$ but two nodes in $\bar{G}$ are connected if and only if they are not connected in $G$. An empty or **trivial graph** is a graph with no links. It is denoted as $\bar{K}_n$ as it is the complement of the complete graph.

A graph is **bipartite** if its nodes can be split into two disjoint (non-empty) subsets $V_1 \subset V$ ($V_1 \neq \phi$) and $V_2 \subset V$ ($V_2 \neq \phi$) and $V_1 \cup V_2 = V$, such that each edge joins a node in $V_1$ and a node in $V_2$. Bipartite graphs do not contain cycles of odd length. If all nodes in $V_1$ are connected to all nodes in $V_2$ the graph is known as complete bipartite graph and denoted by $K_{n_1,n_2}$, where $n_1 = |V_1|$ and $n_2 = |V_2|$ are the number of nodes in $V_1$ and $V_2$, respectively. Finally, a graph is **planar** if it can be drawn in a plane in such a way that no two edges intersect except at a node with which they are both incident.

# 2 Graphs in condensed matter physics

## 2.1 Tight-binding models

In condensed matter physics it is usual to describe solid state and molecular systems by considering the interaction between $N$ electrons whose behavior is determined by a Hamiltonian of the following form:

$$\mathbf{H} = \sum_{n=1}^{N}\left[-\frac{\hbar^2 \nabla_n^2}{2m} + U(r_n) + \frac{1}{2}\sum_{m \neq n} V(r_n - r_m)\right], \quad (2.1)$$

where $U(r_n)$ is an external potential and $V(r_n - r_m)$ is the potential describing the interactions between electrons. Using the second quantization formalism of quantum mechanics this Hamiltonian can be written as:

$$\hat{H} = -\sum_{ij} t_{ij} \hat{c}_i^\dagger \hat{c}_j + \frac{1}{2}\sum_{ijkl} V_{ijkl} \hat{c}_i^\dagger \hat{c}_k^\dagger \hat{c}_l \hat{c}_j, \quad (2.2)$$

where $\hat{c}_i^\dagger$ and $\hat{c}_i$ are 'ladder operators', $t_{ij}$ and $V_{ijkl}$ are integrals which control the hopping of an electron from one site to another and the interaction between electrons, respectively. They are usually calculated directly from finite basis sets (Canadell et al., 2012).

In the tight-binding approach for studying solids and certain classes of molecules, the interaction between electrons is neglected and $V_{ijkl} = 0, \forall i,j,k,l$. This method, which is known as the Hückel molecular orbital method in chemistry, can be seen as very drastic in its approximation, but let us think of the physical picture behind it (Kutzelnigg, 2006; Powell, 2009). We concentrate our discussion on alternant conjugated molecules in which single and double bonds alternate. Consider a molecule like benzene in which every carbon atom has an $sp_2$ hybridization. The frontal overlapping $sp_2 - sp_2$ of adjacent carbon atoms creates very stable $\sigma$-bonds, while the lateral overlapping $p - p$ between adjacent carbon atoms creates very labile $\pi$-bonds. Thus it is clear from the reactivity of this molecule that a $\sigma - \pi$ separation is plausible and we can consider that our basis set consists of orbitals centred on the particular carbon atoms in such a way that there is only one orbital per spin state at each site. Then we can write the Hamiltonian of the system as:

$$\hat{H}_{tb} = -\sum_{ij} t_{ij} \hat{c}_{i\rho}^\dagger \hat{c}_{i\rho}, \quad (2.3)$$

where $\hat{c}_{i\rho}^{(\dagger)}$ creates (annihilates) an electron with spin $\rho$ in a $\pi$ (or other) orbital centred at the atom $i$. We can now separate the in-site energy $\alpha_i$ from the transfer energy $\beta_{ij}$ and write the Hamiltonian as

$$\hat{H}_{tb} = \sum_{ij} \alpha_i \hat{c}_{i\rho}^\dagger \hat{c}_{i\rho} + \sum_{\langle ij \rangle \rho} \beta_{ij} \hat{c}_{i\rho}^\dagger \hat{c}_{i\rho}, \quad (2.4)$$

where the second sum is carried out over all pairs of nearest-neighbors. Consequently, in a molecule or solid with $N$ atoms the Hamiltonian (2.3) is reduced to an $N \times N$ matrix,

$$H_{ij} = \begin{cases} \alpha_i & \text{if } i = j \\ \beta_{ij} & \text{if } i \text{ is connected to } j \\ 0 & \text{otherwise.} \end{cases} \quad (2.5)$$

Due to the homogeneous geometrical and electronic configuration of many systems analyzed by this method we may take $\alpha_i = \alpha, \forall i$ (Fermi energy) and $\beta_{ij} = \beta \approx -2.70 eV$ for all pairs of connected atoms. Thus,

$$\mathbf{H} = \alpha \mathbf{I} + \beta \mathbf{A}, \quad (2.6)$$

where $I$ is the identity matrix, and $\mathbf{A}$ is the adjacency matrix of the graph representing the carbon-skeleton of the molecule. The Hamiltonian and the adjacency matrix of the graph have the same eigenfunctions $\varphi_j$ and their eigenvalues are simply related by:

$$\mathbf{H}\varphi_j = E_j \mathbf{A}, \quad \mathbf{A}\varphi_j = \lambda_j \mathbf{H}, \quad E_j = \alpha + \beta \lambda_j. \quad (2.7)$$

Hence everything we have to do in the analysis of the electronic structure of molecules or solids that can be represented by a tight-binding Hamiltonian, is to study the spectra of the graphs associated with them. The study of spectral properties of graphs represents an entire area of research in algebraic graph theory. The spectrum of a matrix is the set of eigenvalues of the matrix together with their multiplicities. For the case of the adjacency matrix let $\lambda_1(\mathbf{A}) \geq \lambda_2(\mathbf{A}) \geq \cdots \geq \lambda_n(\mathbf{A})$ be the distinct eigenvalues of $\mathbf{A}$ and let $m(\lambda_1(\mathbf{A})), m(\lambda_2(\mathbf{A})), \cdots, m(\lambda_n(\mathbf{A}))$ be their algebraic multiplicities, i.e., the number of times each of them appears as an eigenvalue of $\mathbf{A}$. Then the spectrum of $\mathbf{A}$ can be written as

$$Sp\mathbf{A} = \begin{pmatrix} \lambda_1(\mathbf{A}) & \lambda_2(\mathbf{A}) & \cdots & \lambda_n(\mathbf{A}) \\ m(\lambda_1(\mathbf{A})) & m(\lambda_2(\mathbf{A})) & \cdots & m(\lambda_n(\mathbf{A})) \end{pmatrix}. \quad (2.8)$$

The total $\pi$ (molecular) energy is given by

$$E = \alpha n_e + \beta \sum_{j=1}^{n} g_j \lambda_j, \quad (2.9)$$

where $n_e$ is the number of $\pi$-electrons in the molecule and $g_j$ is the occupation number of the $j$-th molecular orbital. For neutral conjugated systems in their ground state we have (Gutman, 2005),

$$E = \begin{cases} 2 \sum_{j=1}^{n/2} \lambda_j & n \text{ even,} \\ 2 \sum_{j=1}^{(n+1)/2} \lambda_j + \lambda_{(j+1)/2} & n \text{ odd.} \end{cases} \quad (2.10)$$

Because an alternant conjugated hydrocarbon has a bipartite molecular graph: $\lambda_j = -\lambda_{n-j+1}$ for all $j = 1, 2, \ldots, n$. In a few molecular systems the spectrum of the adjacency matrix is known. For instance (Kutzelnigg, 2006), we have
   i)    Polyenes $C_nH_{n+2}$

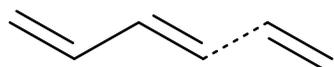

$$\lambda_j(\mathbf{A}) = 2\cos\left(\frac{\pi j}{n+1}\right), \; j = 1, \ldots, n, \quad (2.11)$$

   ii)    Cyclic polyenes $C_nH_n$

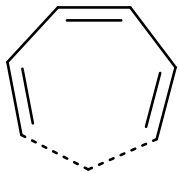

$$\lambda_j(\mathbf{A}) = 2\cos\left(\frac{2\pi j}{n}\right), \ j=1,\ldots,n, \ \lambda_j = \lambda_{n-j} \tag{2.12}$$

iii) Polyacenes,

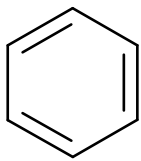    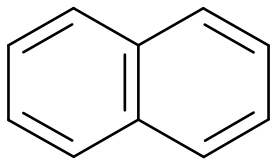    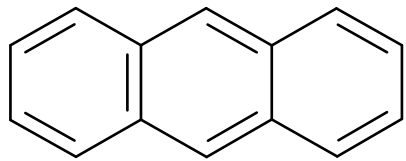

$N=1$ $\qquad\qquad\qquad\qquad N=2 \qquad\qquad\qquad\qquad N=3$

$$\lambda_r(\mathbf{A}) = 1; \lambda_s(\mathbf{A}) = -1;$$
$$\lambda_k(\mathbf{A}) = \pm\frac{1}{2}\left\{1 \pm 9 + 8\cos\frac{k\pi}{N+1}\right\}, k=1,\ldots,N \tag{2.13}$$

A few bounds exist for the total energy of systems represented by graphs with $n$ vertices and $m$ edges. For instance,

$$\sqrt{2m + n(n-1)(\det \mathbf{A})^{n/2}} \le E \le \sqrt{mn} \tag{2.14}$$

and if $G$ is a bipartite graph with $n$ vertices and $m$ edges then,

$$E \le 4m/n + \sqrt{(n-2)(2m - 8m^2/n^2)}. \tag{2.15}$$

### 2.1.1 Nullity and zero-energy states

Another characteristic of a graph which is related to an important molecular property is the nullity. The **nullity** of a graph, denoted by $\eta = \eta(G)$, is the algebraic multiplicity of the zero eigenvalue in the spectrum of the adjacency matrix of the graph (Borovićanin, Gutman, 2009). This property is very relevant for the stability of alternant unsaturated conjugated hydrocarbons. An alternant unsaturated conjugated hydrocarbon with $\eta = 0$ is predicted to have a closed-shell electron configuration. Otherwise, the respective molecule is predicted to have an open-shell electron configuration. That is, when $\eta > 0$ the molecule has unpaired electrons in the form of radicals which are relevant for several electronic and magnetic properties of materials. In a molecule with an even number of atoms, $\eta$ is either zero or it is an even positive integer.

A few important facts about the nullity of graphs are the following. Let $M = M(G)$ be the size of the maximum matching of a graph, i.e., the maximum number of mutually non-adjacent edges of $G$. Let $T$ be a tree with $n \ge 1$ vertices. Then,

$$\eta(T) = n - 2M. \tag{2.16}$$

If $G$ is a bipartite graph with $n \geq 1$ vertices and no cycle of length $4s$ ($s = 1, 2, \ldots$), then

$$\eta(G) = n - 2M. \tag{2.17}$$

Also for a bipartite graph $G$ with incidence matrix $\nabla$, $\eta(G) = n - 2r(\nabla)$, where $r(\nabla)$ is the rank of $\nabla = \nabla(G)$. In the particular case of of benzenoid graphs $Bz$, which may contain cycles of length $4s$, the nullity is given by

$$\eta(Bz) = n - 2M. \tag{2.18}$$

Some known bounds for the nullity of graphs are the following (Cheng, Liu, 2007). Let $G$ be a graph with $n$ vertices and at least one cycle,

$$\eta(G) \leq \begin{cases} n - 2g(G) + 2 & g(G) \equiv 0 \pmod{4}, \\ n - 2g(G) & \text{otherwise,} \end{cases} \tag{2.19}$$

where $g(G)$ is the girth of the graph.

If there is a path of length $d(p,q)$ between the vertices $p$ and $q$ of $G$

$$\eta(G) \leq \begin{cases} n - d(p,q) & \text{if } d(p,q) \text{ is even,} \\ n - d(p,q) - 1 & \text{otherwise.} \end{cases} \tag{2.20}$$

Let $G$ be a simple connected graph of diameter $D$. Then

$$\eta(G) \leq \begin{cases} n - D & \text{if } D \text{ is even,} \\ n - D - 1 & \text{otherwise.} \end{cases} \tag{2.21}$$

## 2.2 Hubbard model

Let us now consider one of the most important models in theoretical physics: the Hubbard model. This model accounts for the quantum mechanical motion of electrons in a solid or conjugated hydrocarbon and includes non-linear repulsive interactions between electrons. In brief, the interest in this model is due to the fact that it exhibits various interesting phenomena including metal–insulator transition, antiferromagnetism, ferrimagnetism, ferromagnetism, Tomonaga–Luttinger liquid, and superconductivity (Takasi, 1999).

The Hubbard model can be seen as an extension of the tight-binding Hamiltonian we have studied in the previous section in which we introduce the electron-electron interactions. To keep things simple, we allow onsite interactions only. That is, we consider one orbital per site and $V_{ijkl} \neq 0$ in (2.2) if and only if $i$, $j$, $k$ and $l$ all refer to the same orbital. In this case the Hamiltonian is written as:

$$\mathbf{H} = -t \sum_{i,j,\sigma} A_{ij} \hat{c}_{i\sigma}^\dagger \hat{c}_{j\sigma} + U \sum_i \hat{c}_{i\uparrow}^\dagger \hat{c}_{i\uparrow} \hat{c}_{i\downarrow}^\dagger \hat{c}_{i\downarrow}, \tag{2.22}$$

where $t$ is the hopping parameter and $U > 0$ indicates that the electrons repel each other.

Notice that if there is not electron-electron repulsion ($U = 0$), we recover the tight-binding Hamiltonian studied in the previous section. Thus, in that case all the results given in the previous section are valid for the Hubbard model without interactions. In the case of non-hopping systems, $t = 0$ and the Hamiltonian is reduced to the electron interaction part only. In this case the remaining Hamiltonian is already in a diagonal form and the eigenstates can be easily obtained. The main difficulty arises when both terms are present in the Hamiltonian. However, in **half-filled systems**, the model has nice properties from a mathematical point of view and a few important results have been proved. These systems have attracted a lot of attention after the discovery of graphene. A system is a half-filled one if the number of electrons is the same as the number of sites. That is, because the total number of electrons can be $2n$, these systems have only a half of the maximum number of electrons allowed. This is particularly the case of graphene and other conjugated aromatic systems. Due to the $\sigma - \pi$ separation which we have seen in the previous section, these systems can be considered as half-filled in which each carbon atom provides one $\pi$ electron.

A fundamental result in the theory of half-filled systems is the theorem proved by Lieb (1989). **Lieb's theorem** for repulsive Hubbard model states the following. Let $G = (V, E)$ be a bipartite connected graph representing a Hubbard model, such that $|V| = n$ is even and the nodes of the graph are partitioned into two disjoint subsets $V_1$ and $V_2$. We assume that the hopping parameters are non-vanishing and that $U > 0$. Then the ground states of the model are non-degenerate apart from the trivial spin degeneracy, and have total spin $S_{tot} = ||V_1| - |V_2||/2$.

In order to illustrate the consequences of Lieb's theorem, let us consider two benzenoid systems which can represent graphene nanoflakes. The first of them is realized in the polycyclic aromatic hydrocarbon known as pyrene and it is illustrated in Figure 2.1 (left). The second is a hypothetical graphene nanoflake known as triangulene and is illustrated in Figure 2.1 (right). In both cases we have divided the bipartite graphs into two subsets, one marked by empty circles which corresponds to $V_1$ and the unmarked nodes form the set $V_2$. In the structure of pyrene we can easily check that $|V_1| = |V_2| = 8$ so that the total spin according to Lieb's theorem is $S_{tot} = 0$. Also according to the formula (2.18) given in the previous section pyerene has no zero-energy levels as its nullity is zero, i.e., $\eta(Bz) = 0$. In this case the mean-field Hubbard model solution for this structure reveals no magnetism.

In the case of triangulene it can be seen that $|V_1| = 12$ and $|V_2| = 10$, which gives a total spin $S_{tot} = 1$. Also the nullity of this graph is equal to 2, indicating that it has two zero-energy states. The result given by Lieb's theorem indicates that triangulene has a spin-triplet ground state which means that it has a magnetic moment of $2\mu_B$ per molecule. Thus triangulene and more $\pi$-extended analogues have intramolecular ferromagnetic interactions owing to $\pi$-spin topological structures. Analogues of this molecule have been already obtained in the laboratory (Morita et al., 2011).

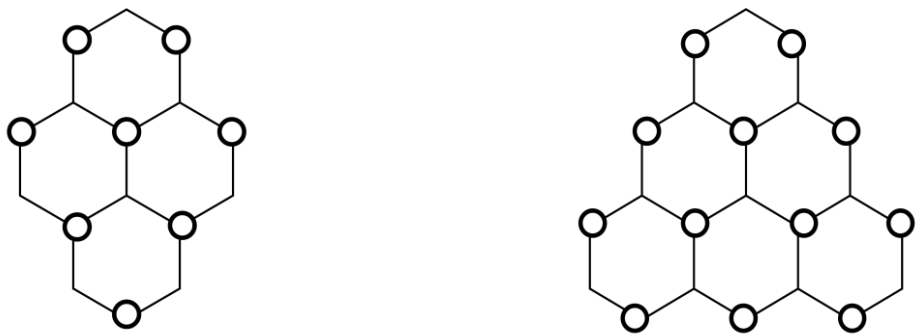

Figure 2.1: Representation of two graphene nanoflakes with closed (left) and open-shell (right) electronic configurations.

# 3 Graphs in statistical physics

The connections between statistical physics and graph theory are extensive and have a long history. A survey on these connections was published already in 1971 by Essam (Essam, 1971); it mainly deals with the Ising model. In the Ising model we consider a set of particles or 'spins', which can be in one of two states. The state of the $i$-th particle is described by the variable $\sigma_i$ which takes one of the two values $\pm 1$. The connection with graph theory comes from the calculation of the partition function of the model. In this chapter we consider that the best way of introducing this connection is through a generalization of the Ising model, the Potts model (Beaudin et al, 2010; Welsh, Merino, 2000).

The Potts model is one of the most important models in statistical physics. In this model we consider a graph $G=(V,E)$ with each node of which we associate a spin. The spin can have one of $q$ values. The basic physical principle of the model is that the energy between two interacting spins is set to zero for identical spins and it is equal to a constant if they are not. A remarkable property of the Potts model is that for $q=3,4$ it exhibits a continuous phase transition between high and low temperature phases. In this case the critical singularities in thermodynamic functions are different from those obtained by using the Ising model. The Potts model has found innumerable applications in statistical physics, e.g. in the theory of phase transitions and critical phenomena, but also outside this context in areas such as magnetism, tumor migration, foam behavior and social sciences.

In the simplest formulation of the Potts model with $q$ states $\{1,2,\cdots,q\}$, the Hamiltonian of the system can have any of the two following forms:

$$\mathbf{H}_1(\omega) = -J \sum_{(i,j)\in E} \delta(\sigma_i, \sigma_j), \tag{3.1}$$

$$\mathbf{H}_2(\omega) = J \sum_{(i,j)\in E} \left[1 - \delta(\sigma_i, \sigma_j)\right], \tag{3.2}$$

where $\omega$ is a configuration of the graph, i.e. an assignment of a spin to each node of $G=(V,E)$; $\sigma_i$ is the spin at node $i$ and $\delta$ is the Kronecker symbol. The model is called ferromagnetic if $J>0$ and antiferromagnetic if $J<0$. We notice here that the Ising model with zero external field is a special case with $q=2$, so that the spins are $+1$ and $-1$.

The probability $p(\omega,\beta)$ of finding the graph in a particular configuration (state) $\omega$ at a given temperature is obtained by considering a Boltzmann distribution and it is given by

$$p(\omega,\beta) = \frac{\exp(-\beta \mathbf{H}_i(\omega))}{Z_i(G)}, \tag{3.3}$$

where $Z_i(G)$ is the partition function for a given Hamiltonian in the Potts model. That is,

$$Z_i(G) = \sum_\omega \exp(-\beta \mathbf{H}_i(\omega)), \tag{3.4}$$

where the sum is over all configurations (states) and $\mathbf{H}_i$ may be either $\mathbf{H}_1$ or $\mathbf{H}_2$. Here $\beta = (k_B T)^{-1}$, where $T$ is the absolute temperature of the system, and $k_B$ is the Boltzmann constant.

For instance, let us consider all the different spin configurations for a cyclic graph with $n = 4$ as given in Figure 3.1. Be aware that there are 4 equivalent configurations for $\omega_2$, $\omega_4$ and $\omega_5$ as well as 2 equivalent configurations for $\omega_3$. The Hamiltonians $\mathbf{H}_1(\omega)$ for these configurations are:

$\mathbf{H}_1(\omega_1) = -4J$; $\mathbf{H}_1(\omega_2) = -2J$; $\mathbf{H}_1(\omega_3) = 0$; $\mathbf{H}_1(\omega_4) = -2J$; $\mathbf{H}_1(\omega_5) = -2J$; $\mathbf{H}_1(\omega_6) = -4J$.

Then, the partition function of the Potts model for this graph is:

$$Z_1(G) = 12\exp(2\beta J) + 2\exp(4\beta J) + 2. \tag{3.5}$$

It is usual to set $K = \beta J$. The probability of finding the graph in the configuration $\omega_2$ is

$$p(\omega_2, \beta) = \frac{\exp(2K)}{12\exp(2K) + 2\exp(4K) + 2}. \tag{3.6}$$

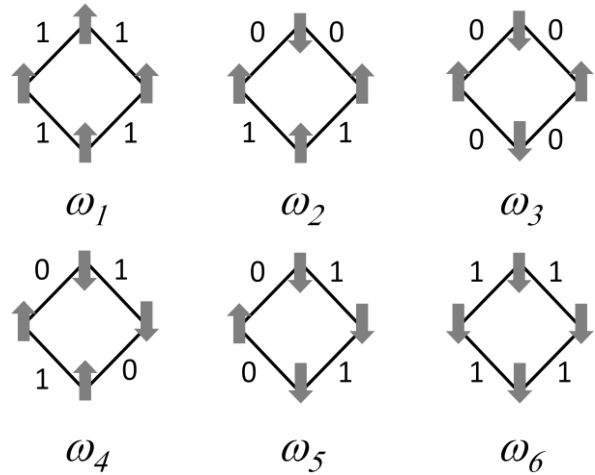

Figure 3.1: Representation of spin configurations in a cycle with four nodes.

The important connection between the Potts model and graph theory comes through the equivalence of this physical model and the graph theoretic concept of the Tutte polynomial. That is, the partition functions of the Potts model can be obtained in the following form:

$$Z_1(G, q, \beta) = q^{k(G)} v^{n-k(G)} T(G; x, y), \tag{3.7}$$

$$Z_2(G, q, \beta) = \exp(-mK) Z_1(G, q, \beta), \tag{3.8}$$

where $q$ is the number of spins in the system, $k(G)$ is the number of connected components of the graph, $v = \exp(K) - 1$, $n$ and $m$ are the number of nodes and edges in the graph, respectively, and $T(G; x, y)$ is the Tutte polynomial, where $x = (q + v)/v$ and $y = \exp(K)$. Proofs of the relationship between the Potts partition function and the Tutte polynomial will not be considered here and the interested reader is directed to the literature to find the details (Bollobás, 1998).

Let us define the **Tutte polynomial** (Ellis-Monagan, Merino, 2011; Welsh, 1999). First, we define the following graph operations. The **deletion** of an edge $e$ in the graph $G$, represented by $G - e$, consists of removing the corresponding edge without changing the rest of the graph, i.e. the end nodes of the edge remain in the graph. The other operation is the **edge contraction** denoted by $G/e$, which consists in gluing together the two end nodes of the edge $e$ and then removing $e$. Both operations, edge deletion and contraction, are commutative, and the operations $G-S$ and $G/S$, where $S$ is a subset of edges, are well defined. We notice here that the graphs created by these transformations are no longer simple graphs, they are pseudographs which may contain self-loops and multiple edges. Let us also define the following types of edges: a **bridge** is an edge whose removal disconnects the graph. A (self) **loop** is an edge having the two end points incident at the same node. Let us denote by $B$ and $L$ the sets of edges which are bridges or loops in the graph.

Then the Tutte polynomial $T(G;x,y)$ is defined by the following recursive formulae:

i) $T(G;x,y) = T(G-e;x,y) + T(G/e;x,y)$ if $e \notin B, L$;

ii) $T(G;x,y) = x^i y^j$ if $e \in B, L$,

where the exponents $i$ and $j$ represent the number of bridges and self-loops in the subgraph, respectively.

Using this definition, we can obtain the Tutte polynomial for the cyclic graph with 4 nodes $C_4$, as illustrated in the Figure 3.2. That is, the Tutte polynomial for $C_4$ is $T(C_4;x,y) = x^3 + x^2 + x + y$. We can substitute this expression into (3.7) to obtain the partition function for the Potts model of this graph,

$$Z_1(G; 2, \beta) = 2^{k(G)} v^{n-k(G)} \left[ \left(\frac{q+v}{v}\right)^3 + \left(\frac{q+v}{v}\right)^2 + \left(\frac{q+v}{v}\right) + 1 + v \right], \quad (3.9)$$

and so we obtain $Z_1(G; 2, \beta) = 12\exp(2K) + 2\exp(4K) + 2$.

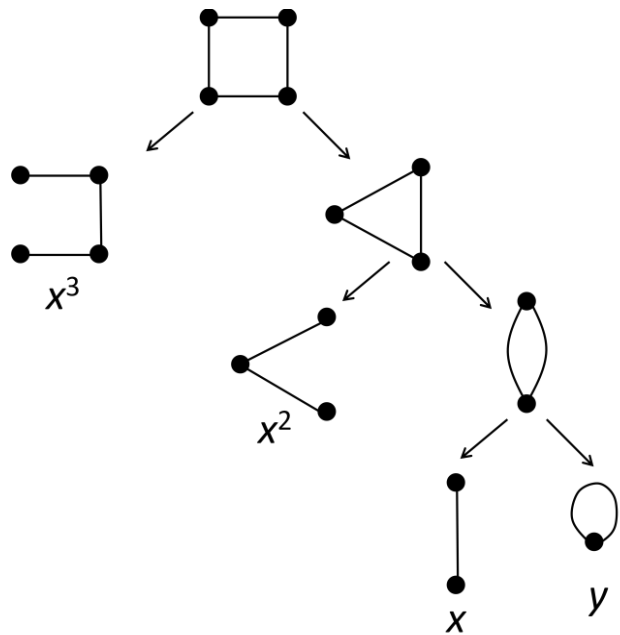

Figure 3.2: Edge deletion and contraction in a cyclic graph with four nodes.

The following is an important mathematical result related to the universality of the Tutte polynomial (Ellis-Monagan, Merino, 2011; Welsh, 1999). Let $f(G)$ be a function on graphs having the following properties:

i)       $f(G) = 1$ if $|V| = 1$ and $|E| = 0$

ii)      $f(G) = af(G-e) + bf(G/e)$ if $e \notin B, L$,

iii)      $f(G \cup H) = f(G)f(H)$; $f(G*H) = f(G)f(H)$, where $G*H$ means that $G$ and $H$ shares at most one node.

Then $f(G)$ is an evaluation of the Tutte polynomial, meaning that it is equivalent to the Tutte polynomial with some specific values for the parameters, and takes the form

$$f(G) = a^{m-n+k(G)} b^{n-k(G)} T\left(G; \frac{f(K_2)}{b}, \frac{f(L)}{a}\right), \qquad (3.10)$$

where $L$ is the graph consisting of a single node with one loop attached, $K_2$ is the complete graph with two nodes.

More formally, the Tutte polynomial is a **generalized Tutte-Gröthendieck** (T-G for short) invariant. To define the T-G invariant, we need the following concepts. Let $S$ and $S'$ be two disjoint subsets of edges. A minor of $G$ is a graph $H$ which is isomorphic to $(G-S)/S'$. Let $\Gamma$ be a class of graphs such that if $G$ is in $\Gamma$ then any minor of $G$ is also in the class. This class is known as minor closed. A graph invariant is a function $f$ on the class of all graphs such that if $G$ and $H$ are isomorphic, then $f(G) = f(H)$. Then, a T-G invariant is a graph invariant $f$ from $\Gamma$ to a commutative ring $\Re$ with unity, such as the conditions (i)-(iii) above are fulfilled. A graph invariant is a function $f$ on the class of all graphs such that $f(G_1) = f(G_2)$ whenever the graphs $G_1$ and $G_2$ are isomorphic. For more details the reader is referred to the specialized literature on this topic.

Some interesting evaluations of the Tutte polynomial are the following:

| | |
|---|---|
| $T(G;1,1)$ | Number of spanning trees of the graph $G$ |
| $T(G;2,1)$ | Number of spanning forests |
| $T(G;1,2)$ | Number of spanning connected subgraphs |
| $T(G;2,2)$ | $2^{|E|}$ |

Let us now consider a proper coloring of a graph $G$, which is an assignment of a color to each node of $G$ such that any two adjacent nodes have different colors. The **chromatic polynomial** $\chi(G;q)$ of the graph $G$ is the number of ways in which $q$ colors can be assigned to the nodes of $G$ such that no two adjacent nodes have the same color. The following are two interesting characteristics of the chromatic polynomial:

i)      $\chi(G;q) = \chi(G-e;q) - \chi(G/e;q)$,

ii)      $\chi(G;q) = q^n$ for the trivial graph on $n$ nodes.

Thus, the chromatic polynomial fulfills the same contraction/deletion rules as the Tutte polynomial. Indeed, the chromatic polynomial is an evaluation of the Tutte polynomial,

$$\chi(G;q) = q^{k(G)} (-1)^{n-k(G)} T(G; 1-q, 0). \qquad (3.11)$$

To see the connection between the Potts model and the chromatic polynomial, we have to consider the Hamiltonian $H_1(G;\omega)$ in the zero temperature limit, i.e. $T \to 0$ ($\beta \to \infty$). When $\beta \to \infty$, the only spin configurations that contributes to the partition function are the ones in which adjacent spins have different values. Then we have that $Z_1(G;q,\beta) \to 1$ in the antiferromagnetic model ($J < 0$). Thus, $Z_1(G;q,\beta \to \infty)$ counts the number of proper colorings of the graph using $q$ colors. The partition function in the $T = 0$ limit of the Potts model is given by the chromatic polynomial

$$Z_1(G;q,-1) = \chi(G) = (-1)^{k(G)}(-1)^n T(G;1-q,0). \tag{3.12}$$

# 4 Feynman graphs

When studying elementary-particle physics, the calculation of higher-order corrections in perturbative quantum field theory naturally leads to the evaluation of Feynman integrals. Feynman integrals are associated to **Feynman graphs**, which are graphs $G = (V, E)$ with $n$ nodes and $m$ edges and some special characteristics (Bogner, 2010; Bogner, Weinzierl, 2010; Weinzierl, 2010). For instance, the edges play a fundamental role in the Feynman graphs as they represent the different particles, such as fermions (edges with arrows), photons (wavy lines), gluons (curly lines). Scalar particles are represented by simple lines. Let us assign a $D$-dimensional momentum vector $q_j$ and a number representing the mass $m_j$ to the $j$-th edge representing the $j$-th particle, where $D$ is the dimension of the space-time. In the theory of Feynman graphs the nodes with degree one are not represented, leaving the edge without the end node. This edge is named an external edge (they are sometimes called *legs*). The rest of edges are called internal. Also, nodes of degree 2 are omitted as they represent mass insertions. Thus, Feynman graphs contain only nodes of degree $k \geq 3$, which represent the interaction of $k$ particles. At each of these nodes the sum of all momenta flowing into the node equals that of the momenta flowing out of it. As usual the number of basic cycles, here termed loops, is given by the cyclomatic number $l = m - n + C$, where $C$ is the number of connected components of the graph.

Here we will only consider Feynman graphs with scalar propagators and we refer to them as **scalar theories**. In scalar theories, the $D$-dimensional Feynman integral has the form

$$I_G = (\mu^2)^{\nu-lD/2} \int \prod_{r=1}^{l} \frac{d^D k_r}{i\pi^{D/2}} \prod_{j=1}^{n} \frac{1}{(-q_j^2 + m_j^2)^{\nu_j}}, \tag{4.1}$$

where $l$ is the number of loops (basic cycles) in the Feynman diagram, $\mu$ is an arbitrary scale parameter used to make the expressions dimensionless, $\nu_j$ is a positive integer number which gives the power to which the propagator occurs, $\nu = \nu_1 + \ldots + \nu_m$, $k_r$ is the independent loop momentum, $m_j$ is the mass of the $j$th particle and

$$q_j = \sum_{j=1}^{l} \rho_{ij} k_j + \sum_{j=1}^{m} \sigma_{ij} p_j \quad \rho_{ij}, \sigma_{ij} \in \{-1, 0, 1\}, \tag{4.2}$$

represents the momenta flowing through the internal lines.

The correspondence between the Feynman integral and the Feynman graph is as follow. An internal edge represents a propagator of the form

$$\frac{i}{q_j^2 - m_j^2}, \tag{4.3}$$

where by abusing of the notation $q_j^2$ represents the inner product of the momentum vector with itself, i.e. $q_j^2 = q_j \cdot q_j^T$. Notice that this is a relativistic propagator which represents a Greens function for integrations over space and time.
Nodes and external edges have weights equal to one. For each internal momentum not constrained by momentum conservation there is also an integration associated.

Now, in order to compute the integral (4.1), we need to assign a (real or complex) variable $x_j$ to each internal edge, which are known as the Feynman parameters. Then, we need to use the Feynman parameter trick for each propagator and evaluate the integrals over the loop momenta $k_1,\ldots,k_l$. As a consequence, we obtain

$$I_G = \frac{\Gamma(\nu - lD/2)}{\prod_{j=1}^{m}\Gamma(\nu_j)} \int_{x_j \geq 0} \left( \prod_{j=1}^{m} dx_j x_j^{\nu_j -1} \right) \times \delta\left(1 - \sum_{i=1}^{m} x_i\right) \frac{U^{\nu-(l+1)D/2}}{F^{\nu - lD/2}} . \tag{4.4}$$

The real connection with the theory of graphs comes from the two terms $U$ and $F$ which are graph polynomials, known as the first and second Symanzik polynomials (sometimes called Kirchhoff-Symanzik polynomials). We will now specify some methods for obtaining these polynomials in Feynman graphs.

## 4.1 Symanzik polynomials and spanning trees

The first **Symanzik polynomial** can be obtained by considering all spanning trees in the Feynman graph. Let $\tau_1$ be the set of spanning trees in the Feynman graph $G$. Then,

$$U = \sum_{T \in \tau_1} \prod_{e_j \notin T} x_j , \tag{4.5}$$

where $T$ is a spanning tree and $x_j$ is the Feynman parameter associated with edge $e_j$.

In order to obtain the second Symanzik polynomial $F$, we have to consider the set of spanning 2-forest $\tau_2$ in the Feynman graph. A **spanning 2-forest** is a spanning forest formed by only two trees. Then, the elements of $\tau_2$ are denoted by $(T_i, T_j)$. The second Symanzik polynomial is given by

$$F = F_0 + U \sum_{i=1}^{m} x_i \frac{m_i^2}{\mu^2} . \tag{4.6}$$

The term $F_0$ is a polynomial obtained from the sets of spanning 2-forests of $G$ in the following way: let $P_{T_i}$ be the set of external momenta attached to the spanning tree $T_i$, which is part of the spanning 2-forest $(T_i, T_j)$. Let $p_k \cdot p_r$ be the Minkowski scalar product of the two momenta vectors associated with the edges $e_k$ and $e_r$, respectively. Then

$$F_0 = \sum_{(T_i,T_j) \in \tau_2} \left( \prod_{e_k \notin (T_i,T_j)} x_k \right) \left( \sum_{p_k \in P_{T_i}} \sum_{p_r \in P_{T_j}} \frac{p_k \cdot p_r}{\mu^2} \right). \tag{4.7}$$

Let us now show how to obtain the Symanzik polynomials for the simple Feynman graph illustrated in the Figure 4.1. For the sake of simplicity, we take all internal masses to be zero.

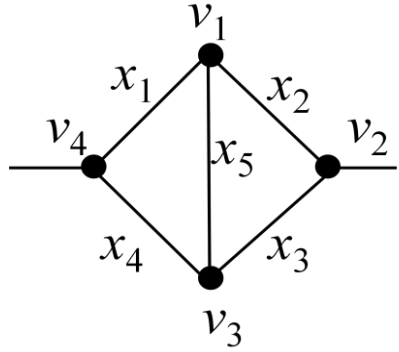

Figure 4.1: Illustration of a Feynman graph with four nodes, five internal and two external edges. The Feynman parameters are represented by $x_i$ on each internal edge.

We first obtain all spanning trees of this graph, which are given in Figure 4.2.

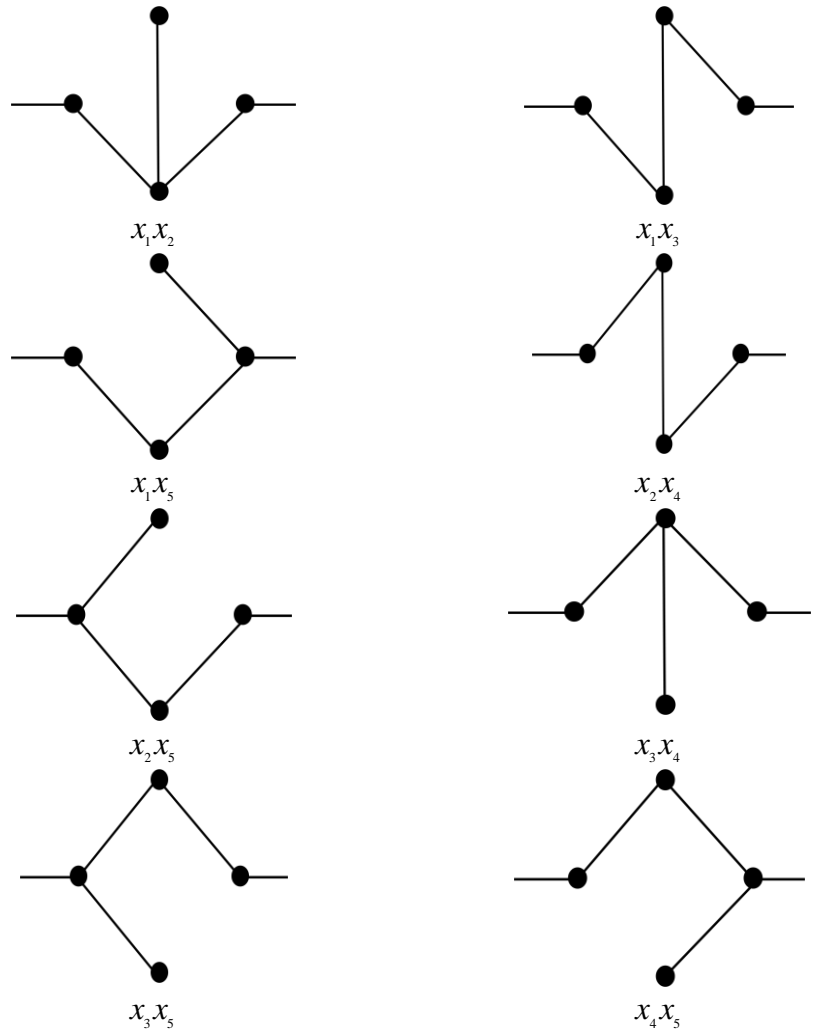

Figure 4.2: Spanning trees of the Feynman graphs represented in Fig. 4.1.

Hence the first Symanzik polynomial is obtained as follows:
$$\begin{aligned} U &= x_1 x_2 + x_1 x_3 + x_1 x_5 + x_2 x_4 + x_2 x_5 + x_3 x_4 + x_3 x_5 + x_4 x_5 \\ &= (x_1 + x_4)(x_2 + x_3) + (x_1 + x_2 + x_3 + x_4) x_5. \end{aligned} \quad (4.8)$$

Now, for the second Symanzik polynomial we obtain all the spanning 2-forests of the graph, which are given in Figure 4.3.

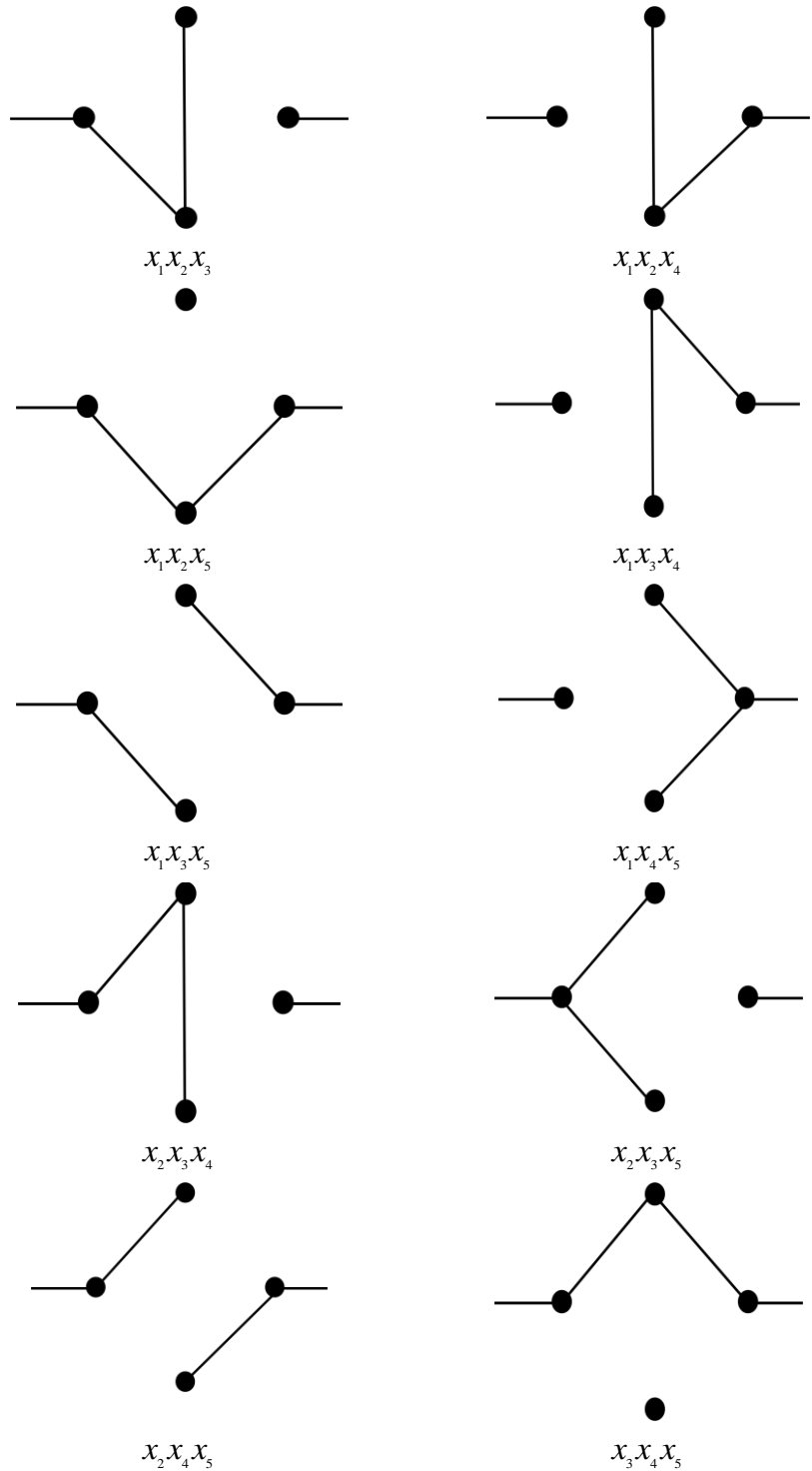

Figure 4.3: Spanning 2-forest of the Feynman graph represented in Fig. 4.1.

We should notice that the terms $x_1x_2x_5$ and $x_3x_4x_5$ do not contribute to $F_0$ because the momentum sum flowing through all cut edges is zero. Thus, we can obtain $F_0$ as follows

$$F = F_0 = \left(x_1 x_2 x_3 + x_1 x_2 x_4 + x_1 x_3 x_4 + x_1 x_3 x_5 + x_1 x_4 x_5 + x_2 x_3 x_4 + x_2 x_3 x_5 + x_2 x_4 x_5\right)\left(\frac{-p^2}{\mu^2}\right)$$

$$= \left[(x_1 + x_2)(x_3 + x_4) x_5 + x_1 x_4 (x_2 + x_3) + x_2 x_3 (x_1 + x_4)\right]\left(\frac{-p^2}{\mu^2}\right). \tag{4.9}$$

## 4.2 Symanzik polynomials and the Laplacian matrix

Another graph-theoretic way of obtaining the Symanzik polynomials is through the use of the Laplacian matrix. The Laplacian matrix for the Feynman graphs is defined as usual for any weighted graph. For instance, for the Feynman graph given in Figure 4.1, the Laplacian matrix is

$$\mathbf{L} = \begin{pmatrix} x_1 + x_2 + x_5 & -x_2 & -x_5 & -x_1 \\ -x_2 & x_2 + x_3 & -x_3 & 0 \\ -x_5 & -x_3 & x_3 + x_4 + x_5 & -x_4 \\ -x_1 & 0 & -x_4 & x_1 + x_4 \end{pmatrix}. \tag{4.10}$$

Then, we can define the auxiliary polynomial $K = \det \mathbf{L}[i]$, where $\mathbf{L}[i]$ denotes the minor of the Laplacian matrix obtained by removing the $i$-th row and column of $\mathbf{L}$. This polynomial is known as the Kirchhoff polynomial of the graph and it is easy to see that it can be defined by

$$K = \sum_{T \in \tau_1} \prod_{e_j \in T} x_j. \tag{4.11}$$

For instance,

$$K = \det \mathbf{L}[1] = \begin{vmatrix} x_2 + x_3 & -x_3 & 0 \\ -x_3 & x_3 + x_4 + x_5 & -x_4 \\ 0 & -x_4 & x_1 + x_4 \end{vmatrix} \tag{4.12}$$

$$= x_1 x_2 x_3 + x_1 x_2 x_4 + x_1 x_2 x_5 + x_1 x_3 x_4 + x_1 x_3 x_5 + x_2 x_3 x_4 + x_2 x_4 x_5 + x_3 x_4 x_5.$$

We transform the Kirchhoff polynomial into the first Symanzik polynomial by setting $U = x_1 \ldots x_m K(x_1^{-1}, \ldots, x_m^{-1})$. That is,

$$U = \frac{x_1 x_2 x_3 x_4 x_5}{x_1 x_2 x_3} + \frac{x_1 x_2 x_3 x_4 x_5}{x_1 x_2 x_4} + \cdots + \frac{x_1 x_2 x_3 x_4 x_5}{x_3 x_4 x_5}$$

$$= x_1 x_2 + x_1 x_3 + x_1 x_5 + x_2 x_4 + x_2 x_5 + x_3 x_4 + x_3 x_5 + x_4 x_5 \tag{4.13}$$

$$= (x_1 + x_4)(x_2 + x_3) + (x_1 + x_2 + x_3 + x_4) x_5.$$

To calculate the second Symanzik polynomial using the Laplacian matrix we have to introduce some modifications. First assign a new parameter $z_j$ to each of the external edges of the Feynman graph. Now, build a diagonal matrix whose diagonal are $D_{ii} = \sum_{j \to i} z_j$, that is the $i$-th diagonal entry of $\mathbf{D}$ represents the sum of the parameters $z_j$ for all the external edges incident with the node $i$. Modify the Laplacian matrix as follows: $\tilde{\mathbf{L}} = \mathbf{L} + \mathbf{D}$. The modified Laplacian matrix $\tilde{\mathbf{L}}$ is the minor of a Laplacian matrix constructed for a modification of the Feynman graph in which all rows and columns corresponding to the external edges are removed (Bogner, Weinzierl, 2010; Weinzierl, 2010). The determinant of the modified Laplacian matrix is

$$W = \det \tilde{\mathbf{L}}, \tag{4.14}$$

and let us expand it in a series of polynomials homogeneous in the variables $z_j$, such that

$$W = W^{(0)} + W^{(1)} + W^{(2)} + \ldots + W^{(t)}, \tag{4.15}$$

where $t$ is the number of external edges. Then the Symanzik polynomials are

$$U = x_1 \ldots x_m W_j^{(1)}\left(x_1^{-1}, \ldots, x_m^{-1}\right) \text{ for any } j,$$

$$F_0 = x_1 \ldots x_m \sum_{(j,k)} \left(\frac{p_j \cdot p_k}{\mu^2}\right) W_{(j,k)}^{(2)}\left(x_1^{-1}, \ldots, x_m^{-1}\right). \tag{4.16}$$

For the Feynman graph given in the previously analyzed example, we have that

$$\tilde{\mathbf{L}} = \begin{pmatrix} x_1 + x_2 + x_5 & -x_2 & -x_5 & -x_1 \\ -x_2 & x_2 + x_3 + z_1 & -x_3 & 0 \\ -x_5 & -x_3 & x_3 + x_4 + x_5 & -x_4 \\ -x_1 & 0 & -x_4 & x_1 + x_4 + z_2 \end{pmatrix}, \tag{4.17}$$

and $W = \det \tilde{\mathbf{L}} = W^{(1)} + W^{(2)}$, where

$$W^{(1)} = (z_1 + z_2)\left(x_1 x_2 x_3 + x_1 x_2 x_4 + x_1 x_3 x_4 + x_2 x_3 x_4 + x_1 x_2 x_5 + x_1 x_3 x_5 + x_2 x_4 x_5 + x_3 x_4 x_5\right), \tag{4.18}$$

$$W^{(2)} = z_1 z_2 \left(x_1 x_3 + x_2 x_3 + x_1 x_4 + x_2 x_4 + x_1 x_5 + x_2 x_5 + x_3 x_5 + x_4 x_5\right). \tag{4.19}$$

With this information the first and second Symanzik polynomials can be easily obtained.

## 4.3 Symanzik polynomials and edge deletion/contraction

The Symanzik polynomials can also be obtained through the graph transformations used to define the Tutte polynomial. That is, the Symanzik polynomials obey the rules for edge deletion and contraction operations which we encountered in the previous section. Recall that the deletion of an edge $e$ in the graph $G$ is represented by $G - e$, and the edge contraction denoted by $G/e$, and that $B$ and $L$ are the sets of edges which are bridges or loops in the graph (see section 3). Then

$$U(G) = U(G/e_j) + x_j U(G - e_j), \tag{4.20}$$

$$F_0(G) = F_0(G/e_j) + x_j F_0(G - e_j), \tag{4.21}$$

for any $e_j \notin B, L$.

Finally, let us mention that there exist factorization theorems for the Symanzik polynomials which are based on a beautiful theorem due to Dodgson (Dodgson, 1866). I cannot resist the temptation to remind the reader that Charles L. Dodgson is better known as Lewis Carroll who has delighted many generations with his *Alice in Wonderland*. These factorization theorems are not given here and the reader is directed to the excellent reviews of Bogner and Weinzierl for details (Bogner, Weinzierl, 2010; Weinzierl, 2010).

## 5 Graphs and electrical networks

The relation between electrical networks and graphs is very natural and is documented in many introductory texts on graph theory. The idea is that a simple **electrical network** can be represented as a graph $G = (V, E)$ in which we place a fixed electrical resistor at each edge of the graph. Therefore, they can also be called **resistor networks**. Let us suppose that we connect a battery across the nodes $u$ and $v$. There are several parameters of an electrical network that can be considered in terms of graph-theoretic concepts but we concentrate here in one which has important connections with other parameters of relevance in physics, namely the effective resistance (Doyle, Snell, 1984). Let us calculate the effective resistance $\Omega(u,v)$ between two nodes by using the Kirchhoff and Ohm laws. For the sake of simplicity we always consider here resistors of 1 Ohm. In the simple case of a tree the effective resistance is simply the sum of the resistances along the path connecting $u$ and $v$. That is, for a tree $\Omega(u,v) = d(u,v)$, where $d(u,v)$ is the shortest path distance between the corresponding nodes (number of links in the shortest path connecting both nodes). However, in the case of two nodes connected by multiple routes, the effective resistance $\Omega(u,v)$ can be obtained by using Kirchhoff's laws. A characteristic of the effective resistance $\Omega(u,v)$ is that it decreases with the increase of the number of routes connecting $u$ and $v$. Thus, in general $\Omega(u,v) \leq d(u,v)$.

An important result about the effective resistance was obtained by Klein and Randić (1993): the effective resistance is a proper distance between the pairs of nodes of a graph. That is,

1. $\Omega(u,v) \geq 0$ for all $u \in V(G), v \in V(G)$.
2. $\Omega(u,v) = 0$ if and only if $u = v$.
3. $\Omega(u,v) = \Omega(v,u)$ for all $u \in V(G), v \in V(G)$..
4. $\Omega(u,w) \leq \Omega(u,v) + \Omega(v,w)$ for all $u \in V(G), v \in V(G), w \in V(G)$.

The **resistance distance** $\Omega(u,v)$ between a pair of nodes $u$ and $v$ in a connected component of a network can be calculated by using the Moore-Penrose generalised inverse $\mathbf{L}^+$ of the graph Laplacian $\mathbf{L}$:

$$\Omega(u,v) = \mathbf{L}^+(u,u) + \mathbf{L}^+(v,v) - 2\mathbf{L}^+(u,v), \tag{5.1}$$

for $u \neq v$.

Another way of computing the resistance distance for a pair of nodes in a network is as follows. Let $\mathbf{L}(G-u)$ be the matrix resulting from removing the $u$th row and column of the Laplacian and let $\mathbf{L}(G-u-v)$ be the matrix resulting from removing both the $u$th and $v$th rows and columns of $\mathbf{L}$. Then it has been proved (Bapat *et al.*, 2003) that

$$\Omega(u,v) = \frac{\det \mathbf{L}(G-u-v)}{\det \mathbf{L}(G-u)}, \tag{5.2}$$

Notice that $\det \mathbf{L}(G-u)$ is the Kirchhoff (Symanzik) polynomial we discussed in the previous section. Yet another way for computing the resistance distance between a pair of nodes in the network is given on the basis of the Laplacian spectra (Xiao, Gutman, 2003)

$$\Omega(u,v) = \sum_{k=2}^{n} \frac{1}{\mu_k} [U_k(u) - U_k(v)]^2, \tag{5.3}$$

where $U_k(u)$ is the $u$-th entry of the $k$-th orthonormal eigenvector associated to the Laplacian eigenvalue $\mu_k$, written in the ordering $0 = \mu_1 < \mu_2 \leq \cdots \leq \mu_n$.

The resistance distance between all pairs of nodes in the network can be represented in the resistance matrix $\boldsymbol{\Omega}$ of the network. This matrix can be written as

$$\boldsymbol{\Omega} = |\mathbf{1}\rangle diag\left\{[\mathbf{L}+(1/n)\mathbf{J}]^{-1}\right\}^T + diag[\mathbf{L}+(1/n)\mathbf{J}]^{-1}\langle\mathbf{1}| - 2(\mathbf{L}+(1/n)\mathbf{J})^{-1}, \tag{5.4}$$

where $\mathbf{J} = |\mathbf{1}\rangle\langle\mathbf{1}|$ is a matrix having all entries equal to 1.

For the case of connected networks the resistance distance matrix can be related to the Moore-Penrose inverse of the Laplacian as shown by Gutman and Xiao (2004):

$$\mathbf{L}^+ = -\frac{1}{2}\left[\mathbf{\Omega} - \frac{1}{n}(\mathbf{\Omega J} + \mathbf{J\Omega}) + \frac{1}{n^2}\mathbf{J\Omega J}\right], \quad (5.5)$$

where $\mathbf{J}$ is as above.

The resistance distance matrix is a matrix of squared *Euclidean distances*. A matrix $\mathbf{M} \in \mathbb{R}^{n \times n}$ is said to be Euclidean if there is a set of vectors $x_1, \ldots, x_n$ such that $M_{ij} = \|x_i - x_j\|^2$. Because it is easy to construct vectors such that $\Omega_{ij} = \|x_i - x_j\|^2$ the resistance distance matrix is squared Euclidean and the resistance distance satisfies the weak triangle inequality

$$\Omega_{ik}^{1/2} \leq \Omega_{ij}^{1/2} + \Omega_{jk}^{1/2}, \quad (5.6)$$

for every pair of nodes in the network.

As we noted in the introduction to this section, effective resistance has connections with other concepts which are of relevance in the applications of mathematics in physics. One of these connections is between the resistance distance and Markov chains. In particular, the resistance distance is proportional to the expected commute time between two nodes for a Markov chain defined by a weighted graph (Ghosh *et al.*, 2008; Doyle and Snell, 1984). If $w_{uv}$ be the weight of the edge $\{u,v\}$, the probability of transition between $u$ and $v$ in the Markov chain defined on the graph is

$$P_{uv} = \frac{w_{uv}}{\sum_{u,v \in E} w_{uv}}. \quad (5.7)$$

The **commuting time** is the time taken by "information" starting at node $u$ to return to it after passing through node $v$. The expected commuting time $\hat{C}_{uv}$ is related to the resistance distance (Ghosh *et al.*, 2008; Doyle and Snell, 1984) by

$$\hat{C}_{uv} = 2(\mathbf{1}^T \mathbf{w})\Omega(u,v), \quad (5.8)$$

where $\mathbf{1}$ is vector of 1s and $\mathbf{w}$ is the vector of link weights. Note that if the network is unweighted $\hat{C}_{uv} = 2m\Omega(u,v)$.

## 6 Graphs and vibrations

In this section we develop some connections between vibrational analysis, which is important in many areas of physics ranging from classical to quantum mechanics, and the spectral theory of graphs. Here we consider the one-dimensional case with a graph $G = (V,E)$ in which every node represents a ball of mass $m$ and every edge represents a spring with the spring constant $m\omega^2$ connecting two balls. The ball-spring network is assumed to be submerged in a thermal bath at temperature $T$. The balls in the graph oscillate under thermal excitation. For the sake of simplicity, we assume that there is no damping and no external forces are applied to the system. Let $x_i$, $i = 1, 2, \cdots, n$ be the coordinates of each node which measures the displacement of the ball $i$ from its equilibrium state $x_i = 0$. For a complete guide to the results to be presented here the reader is directed to Estrada et al. (2012).

## 6.1 Graph vibrational Hamiltonians

Let us start with a Hamiltonian of the oscillator network in the form

$$\mathbf{H}_A = \sum_i \left[ \frac{p_i^2}{2m} + (K - k_i) \frac{m\omega^2 x_i^2}{2} \right] + \frac{m\omega^2}{2} \sum_{\substack{i,j \\ (i<j)}} A_{ij} (x_i - x_j)^2, \quad (6.1)$$

where $k_i$ is the degree of the node $i$ and $K$ is a constant satisfying $K \geq \max_i k_i$. The second term in the right-hand side is the potential energy of the springs connecting the balls, because $x_i - x_j$ is the extension or the contraction of the spring connecting the nodes $i$ and $j$. The first term in the first set of square parentheses is the kinetic energy of the ball $i$, whereas the second term in the first set of square parentheses is a term that avoids the movement of the network as a whole by tying the network to the ground. We add this term because we are only interested in small oscillations around the equilibrium; this will be explained below again.

The Hamiltonian (6.1) can be rewritten as

$$\mathbf{H}_A = \sum_i \left( \frac{p_i^2}{2m} + \frac{Km\omega^2}{2} x_i^2 \right) - \frac{m\omega^2}{2} \sum_{i,j} x_i A_{ij} x_j. \quad (6.2)$$

Let us next consider the Hamiltonian of the oscillator network in the form

$$\mathbf{H}_L = \sum_i \frac{p_i^2}{2m} + \frac{m\omega^2}{2} A_{ij} (x_i - x_j)^2 \quad (6.3)$$

instead of the Hamiltonian $\mathbf{H}_A$ in Eq. (6.2). Because the Hamiltonian $\mathbf{H}_L$ lacks the springs that tie the whole network to the ground (the second term in the first set of parentheses in the right-hand side of Eq. (78), this network can undesirably move as a whole. We will deal with this motion shortly.

The expansion of the Hamiltonian (6.3) as in Eqs. (6.1)-(6.2) now gives

$$\mathbf{H}_L = \sum_i \frac{p_i^2}{2m} + \frac{m\omega^2}{2} \sum_{i,j} x_i L_{ij} x_j, \quad (6.4)$$

where $L_{ij}$ denotes an element of the network Laplacian $\mathbf{L}$.

## 6.2 Network of Classical Oscillators

We start by considering a network of classical harmonic oscillators with the Hamiltonian $H_A$. Here the momenta $p_i$ and the coordinates $x_i$ are independent variables, so that the integration of the factor

$$\prod_i \exp\left[ -\beta \left( \frac{p_i^2}{2m} \right) \right] \quad (6.5)$$

over the momenta $\{p_i\}$ reduces to a constant term, which does not affect the integration over $\{x_i\}$. As a consequence we do not have to consider the kinetic energy and we can write the Hamiltonian in the form

$$\mathbf{H}_A = \frac{m\omega^2}{2} x^T (K\mathbf{I} - \mathbf{A}) x, \tag{6.6}$$

where $x = (x_1, x_2, \cdots, x_n)^T$ and $I$ is the $n \times n$ identity matrix.

The partition function is given by

$$Z = \int e^{-\beta \mathbf{H}_A} \prod_i dx_i = \int dx \exp\left(-\frac{\beta m \omega^2}{2} x^T (K\mathbf{I} - \mathbf{A}) x\right), \tag{6.7}$$

where the integral is an $n$-fold one and can be evaluated by diagonalizing the matrix $\mathbf{A}$. The adjacency matrix can be diagonalized by means of an orthogonal matrix $\mathbf{O}$ as in

$$\mathbf{\Lambda} = \mathbf{O}(K\mathbf{I} - \mathbf{A})\mathbf{O}^T, \tag{6.8}$$

where $\mathbf{\Lambda}$ is the diagonal matrix with eigenvalues $\lambda_\mu$ of $(K\mathbf{I} - \mathbf{A})$ on the diagonal. Let us consider that $K$ is sufficiently large, so that we can make all eigenvalues $\lambda_\mu$ positive. By defining a new set of variables $y_\mu$ by $y = \mathbf{O}x$ and $x = \mathbf{O}^T y$, we can transform the Hamiltonian (6.6) to the form

$$\mathbf{H}_A = \frac{m\omega^2}{2} y^T \mathbf{\Lambda} y = \frac{m\omega_0^2}{2} \sum_\mu y_\mu^2 + \frac{m\omega^2}{2} \sum_\mu \lambda_\mu y_\mu^2. \tag{6.9}$$

Then the integration measure of the $n$-fold integration in Eq. (6.7) is transformed as $\prod_i dx_i = \prod_\mu dy_\mu$, because the Jacobian of the orthogonal matrix $\mathbf{O}$ is unity. Therefore, the multi-fold integration in the partition function (6.7) is decoupled to give

$$Z = \prod_\mu \sqrt{\frac{2\pi}{\beta m \omega^2 \lambda_\mu}}, \tag{6.10}$$

which can be rewritten in terms of the adjacency matrix as

$$Z = \left(\frac{2\pi}{\beta m} \omega^2\right)^{n/2} \frac{1}{\sqrt{\det(K\mathbf{I} - \mathbf{A})}}. \tag{6.11}$$

Since we have made all the eigenvalues of $(K\mathbf{I} - \mathbf{A})$ positive, its determinant is positive.

Now we define an important quantity, the mean displacement of a node from its equilibrium position. It is given by

$$\langle x_p^2 \rangle = \frac{1}{Z} \int x_p^2 e^{-\beta \mathbf{H}_A} \prod_i dx_i, \tag{6.12}$$

which by using the spectral decomposition of $\mathbf{A}$, yields

$$\langle x_p^2 \rangle = \frac{1}{Z} \int \left[\sum_\sigma (\mathbf{O}^T)_{p\sigma} y_\sigma\right]^2 e^{-\beta \mathbf{H}_A} \prod_\mu dy_\mu. \tag{6.13}$$

In the integrand, the odd functions with respect to $y_\mu$ vanish. Therefore, only the terms of $y_\sigma^2$ survive after integration in the expansion of the square parentheses in the integrand. This gives

$$\langle x_p^2 \rangle = \frac{1}{Z} \sum_\sigma O_{\sigma p}^2 \int y_\sigma^2 \exp\left(-\frac{\beta m \omega^2}{2} \lambda_\sigma y_\sigma^2\right) dy_\sigma$$
$$\times \prod_{\mu(\neq \sigma)} \left[\int \exp\left(-\frac{\beta m \omega^2}{2} \lambda_\mu y_\mu^2\right) dy_\mu\right].$$
(6.14)

Comparing this expression with Eq. (6.10), we have

$$\langle x_p^2 \rangle = \frac{1}{\beta m K \omega^2} \left[(\mathbf{I} - \mathbf{A}/K)^{-1}\right]_{pp}.$$
(6.15)

The mean node displacement may be given by the thermal Green's function in the framework of classical mechanics by

$$\langle x_p^2 \rangle = \frac{1}{\beta K m \omega^2} \left[(\mathbf{I} - \mathbf{A}/K)^{-1}\right]_{pq}.$$
(6.16)

This represents a correlation between the node displacements in a network due to small thermal fluctuations.

The same calculation using the Hamiltonian (6.4) gives

$$\langle x_p^2 \rangle' = \frac{1}{\beta m \omega^2} \left(\mathbf{L}^+\right)_{pq}$$
(6.17)

where $\mathbf{L}^+$ is the Moore-Penrose generalized inverse of the Laplacian.

## 6.3 Network of Quantum Oscillators

Here we consider the quantum-mechanical version of the Hamiltonian $\mathbf{H}_A$ in Eq. (6.2) by considering that the momenta $p_j$ and the coordinates $x_i$ are not independent variables. In this case they are operators that satisfy the commutation relation,

$$[x_i, p_j] = i\hbar \delta_{ij}.$$
(6.18)

We use the boson creation and annihilation operators $a_i^\dagger$ and $a_i$ which allow us to write the coordinates and momenta as

$$x_i = \sqrt{\frac{\hbar}{2m\Omega}} \left(a_i^\dagger + a_i\right),$$
(6.19)

$$p_i = \sqrt{\frac{\hbar}{2m\Omega}} \left(a_i^\dagger - a_i\right),$$
(6.20)

where $\Omega = \sqrt{K/m\omega}$. The commutation relation (6.18) yields

$$[a_i, a_j^\dagger] = \delta_{ij}.$$
(6.21)

With the use of these operators, we can recast the Hamiltonian (6.2) into the form

$$\mathbf{H}_A = \sum_i \hbar\Omega\left(a_i^\dagger a_i + \frac{1}{2}\right) - \frac{\hbar\omega^2}{4\Omega}\sum_{i,j}\left(a_i^\dagger + a_i\right)A_{ij}\left(a_j^\dagger + a_j\right). \tag{6.22}$$

Using the spectral decomposition of the adjacency matrix, we generates a new set of boson creation and annihilation operators given by

$$b_\mu = \sum_i O_{\mu i} a_i = \sum_i a_i \left(\mathbf{O}^T\right)_{i\mu}, \tag{6.23}$$

$$b_\mu^\dagger = \sum_i O_{\mu i} a_i^\dagger = \sum_i a_i^\dagger \left(\mathbf{O}^T\right)_{i\mu}, \tag{6.24}$$

Applying the transformations (6.23)-(6.24) to the Hamiltonian (6.22), we can decouple it as

$$\mathbf{H}_A = \sum_\mu \mathbf{H}_\mu, \tag{6.25}$$

with

$$\mathbf{H}_\mu = \hbar\Omega\left[1 + \frac{\omega^2}{2\Omega^2}(\lambda_\mu - K)\right]\left(b_\mu^\dagger b_\mu + \frac{1}{2}\right) + \frac{\hbar\omega^2}{4\Omega}(\lambda_\mu - K)\left[\left(b_\mu^\dagger\right)^2 + \left(b_\mu\right)^2\right]. \tag{6.26}$$

In order to go further, we now introduce an approximation in which each mode of oscillation does not get excited beyond the first excited state. In other words, we restrict ourselves to the space spanned by the ground state (the vacuum) $|\mathrm{vac}\rangle$ and the first excited states $b_\mu^\dagger|\mathrm{vac}\rangle$. Then the second term of the Hamiltonian (6.26) does not contribute and we therefore have

$$\mathbf{H}_\mu = \hbar\Omega\left[1 + \frac{\omega^2}{2\Omega^2}(\lambda_\mu - K)\right]\left(b_\mu^\dagger b_\mu + \frac{1}{2}\right) \tag{6.27}$$

within this approximation. This approximation is justified when the energy level spacing $\hbar\Omega$ is much greater than the energy scale of external disturbances, (specifically the temperature fluctuation $k_B T = 1/\beta$, in assuming the physical metaphor that the system is submerged in a thermal bath at the temperature $T$), as well as than the energy of the network springs $\hbar\omega$, i.e. $\beta\hbar\Omega \gg 1$ and $\Omega \gg \omega$. This happens when the mass of each oscillator is small, when the springs connecting to the ground $m\Omega^2$ are strong, and when the network springs $m\omega^2$ are weak. Then an oscillation of tiny amplitude propagates over the network. We are going to work in this limit hereafter.

We are now in a position to compute the partition function as well as the thermal Green's function quantum-mechanically. As stated above, we consider only the ground state and one excitation from it. Therefore we have the quantum-mechanical partition function in the form

$$Z^A = \langle\mathrm{vac}|e^{-\beta\mathbf{H}_A}|\mathrm{vac}\rangle$$
$$= \prod_\mu \exp\left\{-\frac{\beta\hbar\Omega}{2}\left[1 + \frac{\omega^2}{2\Omega^2}(\lambda_\mu - K)\right]\right\}. \tag{6.28}$$

The diagonal thermal Green's function giving the mean node displacement in the quantum mechanical framework is given by

$$\langle x_p^2\rangle = \frac{1}{Z}\langle\mathrm{vac}|a_p e^{-\beta\mathbf{H}_A} a_p^\dagger|\mathrm{vac}\rangle, \tag{6.29}$$

which indicates how much an excitation at the node $p$ propagates throughout the graph before coming back to the same node and being annihilated. Let us compute the quantity (6.29) by

$$\langle x_p^2 \rangle = e^{-\beta\hbar\Omega} \left( \exp\left[ \frac{\beta\hbar\omega^2}{2\Omega} \mathbf{A} \right] \right)_{pp}, \qquad (6.30)$$

where we have used Eq. (6.8). Similarly, we can compute the off-diagonal thermal Green's function as

$$\langle x_p, x_q \rangle = e^{-\beta\hbar\Omega} \left( \exp\left[ \frac{\beta\hbar\omega^2}{2\Omega} \mathbf{A} \right] \right)_{pq}. \qquad (6.31)$$

The same quantum-mechanical calculation by using the Hamiltonian $H_L$ in Eq. (6.3) gives

$$\langle x_p, x_q \rangle = 1 + \lim_{\Omega \to 0} O_{2p} O_{2q} \exp\left[ -\frac{\beta\hbar\omega^2}{2\Omega} \mu_2 \right], \qquad (6.32)$$

where $\mu_2$ is the second eigenvalue of the Laplacian matrix.

# 7 Random graphs

The study of random graphs is one of the most important areas of theoretical graph theory. Random graphs have found multiple applications in physics and they are used today as a standard null model in simulating many physical processes on graphs and networks. There are several ways of defining a random graph, that is, a graph in which, given a set of nodes, the edges connecting them are selected in a random way. The simplest model of random graph was introduced by Erdös and Rényi (1959). The construction of a random graph in this model starts by considering $n$ isolated nodes. Then, with probability $p > 0$ a pair of nodes is connected by an edge. Consequently, the graph is determined only by the number of nodes and edges such that it can be written as $G(n,m)$ or $G(n,p)$. In Fig. 7 we illustrate some examples of **Erdös-Rényi random graphs** with the same number of nodes and different linking probabilities.

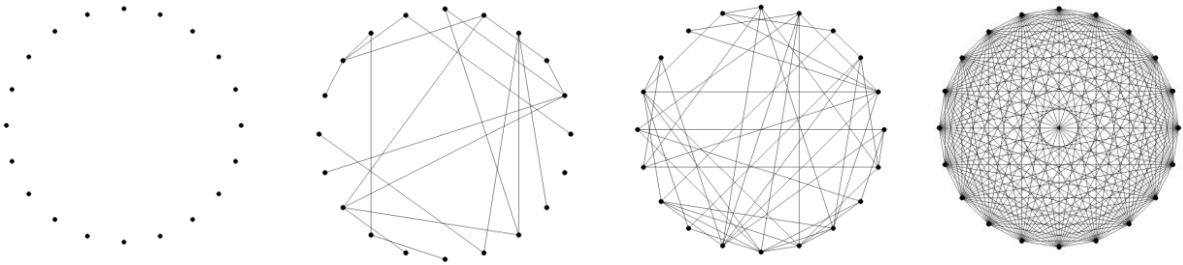

Figure 7.1: Illustration of the changes of an Erdös-Rényi random network with 20 nodes and probabilities that increases from zero (left) to one (right).

A few properties of Erdös-Rényi (ER) random graphs are summarised below.
i) The expected number of edges per node:
$$\bar{m} = \frac{n(n-1)p}{2}. \qquad (7.1)$$

ii) The expected node degree:
$$\bar{k} = (n-1)p.$$

iii) The average path length for large $n$:
$$\bar{l}(H) = \frac{\ln n - \gamma}{\ln(pn)} + \frac{1}{2}, \tag{7.2}$$
where $\gamma \approx 0.577$ is the Euler-Mascheroni constant.

iv) The average clustering coefficient (see Eq. (1.12)):
$$\bar{C} = p = \delta(G). \tag{7.3}$$

v) When increasing $p$, most nodes tends to be clustered in one giant component, while the rest of nodes are isolated in very small components (see Fig. 7.2).

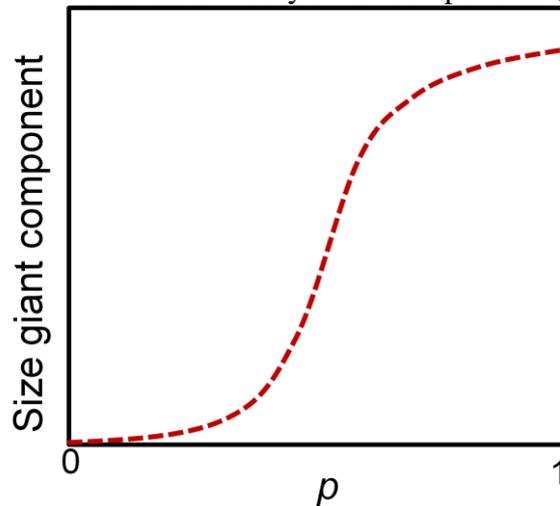

Figure 7.2: Change of the size of the giant connected component in an ER random graph as probability is increased.

vi) The structure of $G_{ER}(n, p)$ changes as a function of $p = \bar{k}/(n-1)$ giving rise to the following three stages (see Fig. 7.3):
  a) *Subcritical* $\bar{k} < 1$, where all components are simple and very small. The size of the largest component is $S = O(\ln n)$.
  b) *Critical* $\bar{k} = 1$, where the size of the largest component is $S = \Theta(n^{2/3})$.
  c) *Supercritical* $\bar{k} > 1$, where the probability that $(f - \varepsilon)n < S < (f + \varepsilon)n$ is 1 when $n \to \infty$ $\varepsilon > 0$ for, where $f = f(\bar{k})$ is the positive solution of the equation: $e^{-\bar{k}f} = 1 - f$. The rest of the components are very small, with the second largest having size about $\ln n$.

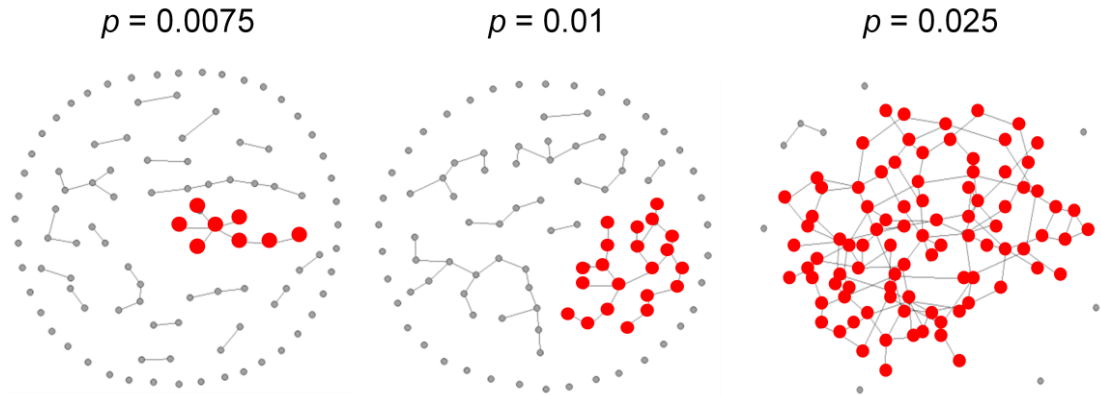

Figure 7.3: Examples of the different stage of the change of an ER random graph with the increase in probability: subcritical (left), critical (centre) and supercritical (right).

vii) The largest eigenvalue of the adjacency matrix in an ER network grows proportionally to $n$ (Janson, 2005): $\lim_{n\to\infty}(\lambda_1(\mathbf{A})/n) = p$.

viii) The second largest eigenvalue grows more slowly than $\lambda_1$: $\lim_{n\to\infty}(\lambda_2(\mathbf{A})/n^\varepsilon) = 0$ for every $\varepsilon > 0.5$.

ix) the smallest eigenvalue also grows with a similar relation to $\lambda_2(\mathbf{A})$: $\lim_{n\to\infty}(\lambda_n(\mathbf{A})/n^\varepsilon) = 0$ for every $\varepsilon > 0.5$.

x) the spectral density of an ER random network follows the **Wigner's semicircle law** (Wigner, 1955), which is simply written as (see Fig. 7.4):

$$\rho(\lambda) = \begin{cases} \dfrac{\sqrt{4-\lambda^2}}{2\pi} & -2 \leq \lambda/r \leq 2, \quad r = \sqrt{np(1-p)} \\ 0 & \text{otherwise.} \end{cases} \quad (7.4)$$

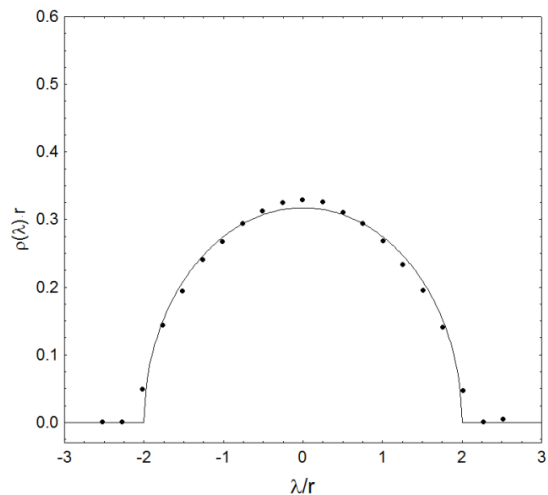

Figure 7.4: Illustration of the Wigner semicircle law for the spectral density of an ER random graph.

## 8 Introducing complex networks

In the rest of this Chapter we are going to study so-called complex networks. Complex networks can be considered as the skeleton of complex systems in a variety of scenarios ranging from social and ecological to biological and technological systems. Their study has become a major field of interdisciplinary research in XXI century with an important participation of physicists who have contributed significantly by creating new models and adapting others known in physics to the study of the topological and dynamical properties of these networks. A number of universal topological properties which explain some of the dynamical and functional properties of networks have been introduced, such as 'small-world' and 'scale-free' phenomena; these will be analyzed briefly in the next sections.

There is much confusion about what a complex network is. To start with we should attempt a clarification about what a complex system is. There is not a clear cut definition of a complex system. First, it must be clear that the concept of complexity is a twofold one: it may refer to a quality of the system or to a quantitative concept. In the first case, complexity is what makes the system complex. In the second, it is a continuum embracing both the simple and the complex according to a given measure of complexity. Standish (2008) has stressed that as a quality "*complexity of a system referes to the presence of emergence in the system, or the exhibition of behaviour not specified in the system specification*". In other words, any complex system "*display organization without any external organizing principle being applied*" (Ottino, 2003). When we speak as complexity as a quantity it "*referes to the amount of information needed to specify the system*".

Then, what is a complex network? Before attempting to answer this question let us try to make a classification of some of the systems represented by networks (see Estrada, 2011 in Further reading) by considering the nature of the links they represent. Some examples of these classes are:

- *Physical linking*: pairs of nodes are physically connected by a *tangible link*, such as a cable, a road, a vein, etc. Examples are: Internet, urban street networks, road networks, vascular networks, etc.
- *Physical interactions*: links between pairs of nodes represents *interactions* which are determined by a *physical force*. Examples are: protein residue networks, protein-protein interaction networks, etc.
- *'Ethereal' connections*: links between pairs of nodes are *intangible*, such that information sent from one node is received at another irrespective of the 'physical' trajectory. Examples are: WWW, airports network.
- *Geographic closeness*: nodes represent regions of a surface and their connections are determined by their *geographic proximity*. Examples are: countries in a map, landscape networks, etc.
- *Mass/energy exchange*: links connecting pairs of nodes indicate that some *energy or mass* has been *transferred* from one node to another. Examples are: reaction networks, metabolic networks, food webs, trade networks, etc.
- *Social connections*: links represent any kind of *social relationship* between nodes. Examples are: friendship, collaboration, etc.
- *Conceptual linking*: links indicate *conceptual relationships* between pairs of nodes. Examples are: dictionaries, citation networks, etc.

Now, let us try to characterize the complexity of these networks by giving the minimum amount of information needed to describe them. For the sake of comparison let us also consider a regular and a random graph of the same size of the real-world networks we want to describe. For the case of a regular graph we only need to specify the number of nodes and the degree of the nodes (recall that every node has the same degree). With this information many nonisomorphic graphs can be constructed, but many of their topological and combinatorial properties are determined by the information provided. In the case of the random network we need to specify the number of nodes and the probability for joining pairs of nodes. As we have seen in the previous section most of the structural properties of these networks are determined by this information. In contrast, to describe the structure of one of the networks representing a real-world system we need an awful amount of information, such as: number of nodes and links, degree distribution, degree-degree correlation, diameter, clustering, presence of communities, patterns of communicability, and other properties that we will study in this section. However, even in this case a complete description of the system is still far away. Thus, the network representation of these systems deserves the title of **complex networks** because their topological structures cannot be trivially described like in the cases of random or regular graphs. In closing, when referring to complex networks we are making implicit allusion to the topological or structural complexity of the graphs representing a complex system. We will consider some general topological and dynamical properties of these networks in the following sections and the reader is recommended to consult the Further Reading section at the end of this Chapter for more details and examples of applications.

## 9 Small-World networks

One of the most popular concepts in network theory is that of the 'small-world'. Practically in every language and culture we have a phrase saying that the World is small enough so that a randomly chose person has a connection with some of our friends. The empirical grounds for this 'concept' come from an experiment carried out by Stanley Milgram in 1967 (Milgram 1967). Milgram asked some randomly selected people in the U.S. cities of Omaha (Nebraska) and Wichita (Kansas) to send a letter to a target person who lives in Boston (Massachusetts) on the East Coast. The rules stipulate that the letter should be sent to somebody the sender knows personally. Despite the senders and the target being separated by about 2000 km the results obtained by Milgram were surprising because:
  i) The average number of steps needed for the letters to arrive to its target was around 6.
  ii) There was a large group inbreeding, which resulted in acquaintances of one individual feedback into his/her own circle, thus usually eliminating new contacts.

The assumption that the underlying social network is a random one with characteristics like the ER network fails to explain these findings. We already know that an Erdös-Rényi random network displays a very small average path length, but it fails in reproducing the large group inbreeding observed because the number of triangles and the clustering coefficient in the ER network are very small. In 1998 Watts and Strogatz (1998) proposed a model which reproduces the two properties mentioned before in a simple way. Let $n$ be the number of nodes and $k$ be an even number, the Watt-Strogatz model starts by using the following construction. Place all nodes in a circle and connect every node to its first $k/2$ clockwise nearest neighbours as well as to its $k/2$ counterclockwise nearest neighbours (see Figure 9.1). This will create a ring, which for $k>2$ is full of triangles and consequently has a large clustering coefficient. The average clustering coefficient for these networks is given by (Barrat and Weigt, 2000)

$$\bar{C} = \frac{3(k-2)}{4(k-1)},$$
(9.1)

which means that $\bar{C} = 0.75$ for very large values of $k$.

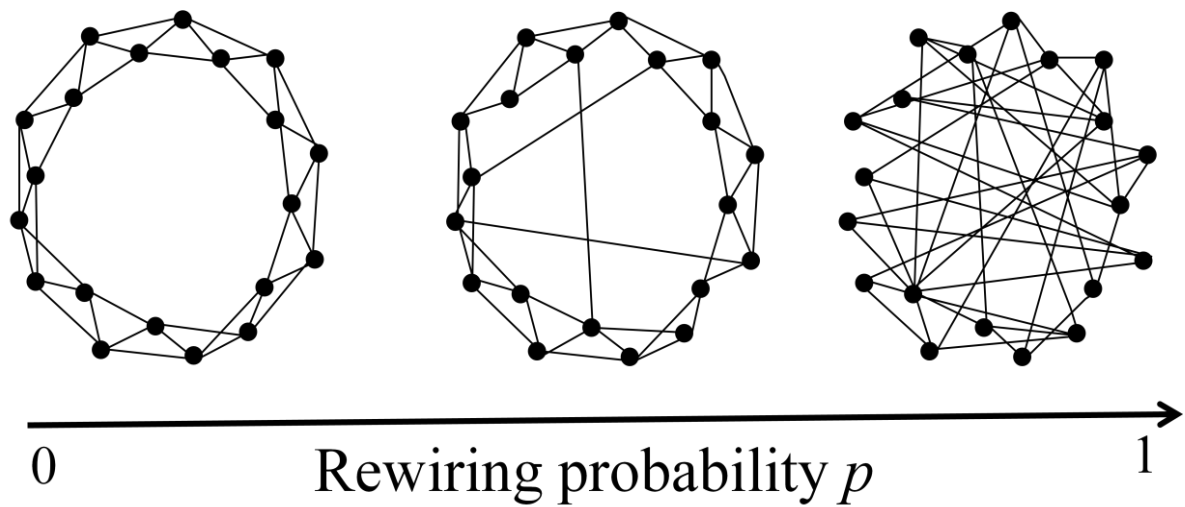

Figure 9.1: Schematic representation of the evolution of the rewiring process in the Watts-Strogatz model.

As can be seen in Fig. 9.1 (top left) the shortest path distance between any pair of nodes which are opposite to each other in the network is relatively large. This distance is, in fact, equal to $\left\lceil \dfrac{n}{k} \right\rceil$. Then

$$\bar{l} \approx \frac{(n-1)(n+k-1)}{2kn}.\tag{9.2}$$

This relatively large average path length is far from that of the Milgram experiment. In order to produce a model with small average path length and still having relatively large clustering, Watts and Strogatz consider a probability for rewiring the links in that ring. This rewiring makes the average path length decreases very fast while the clustering coefficient still remains high. In Fig. 9.2 we illustrate what happens to the clustering and average path length as the rewiring probability change from 0 to 1 in a network.

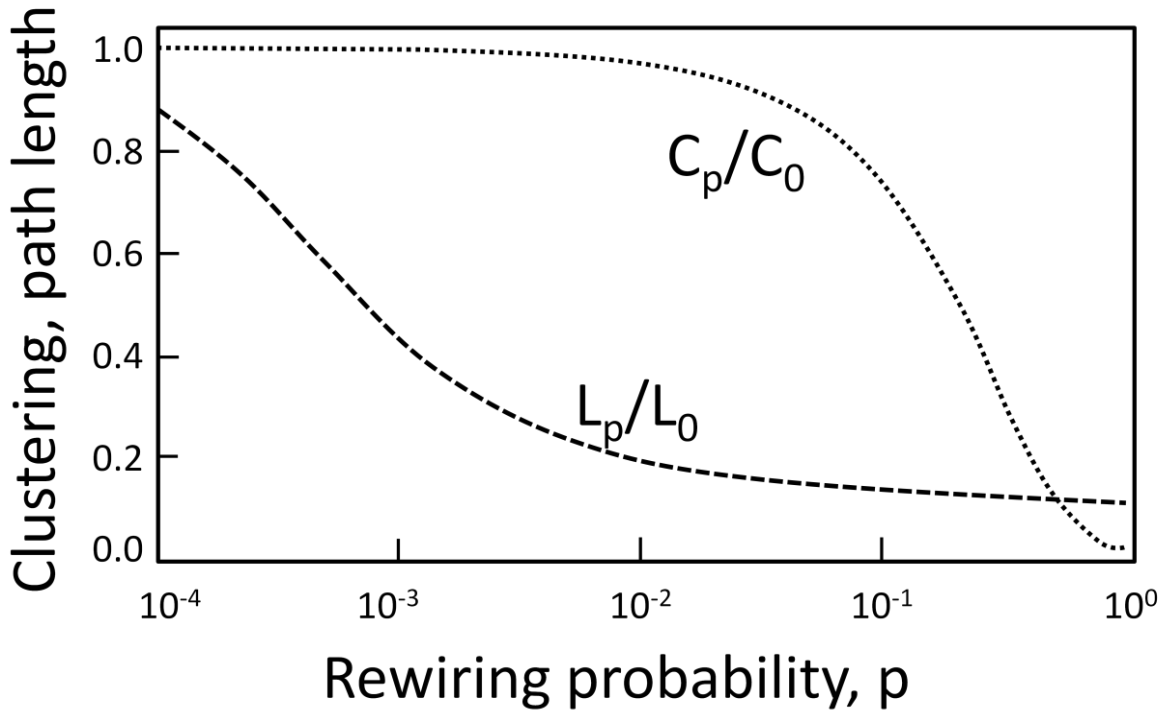

Figure 9.2: Schematic representation of the variation in the average path length and clustering coefficient with the change of the rewiring probability in the Watts-Strogatz model.

## 10 Degree distributions

One of the network characteristics that has received much attention in the literature is the statistical distribution of the node degrees. Let $p(k) = n(k)/n$, where $n(k)$ is the number of nodes having degree $k$ in the network of size $n$. That is, $p(k)$ represents the probability that a node selected uniformly at random has degree $k$. The histogram of $p(k)$ versus $k$ represents the degree distribution for the network. There are hundreds of statistical distributions in which the node degrees of a network can fit. A typical distribution which is expected for a random network of the type of Erdös-Rényi is the Poisson distribution. However, a remarkable characteristic of complex networks is that many of them display some kind of 'fat-tailed' degree distributions. In these distributions a few nodes appear with very large degree while most of the nodes have relatively small degrees. The prototypical example of these distributions is the power-law one, which is illustrated in the Figure 10.1, but others like lognormal, Burr, logGamma, Pareto, etc. (Foss *et al.*, 2011) fall in the same category.

| Poisson distribution | Power-law distribution |

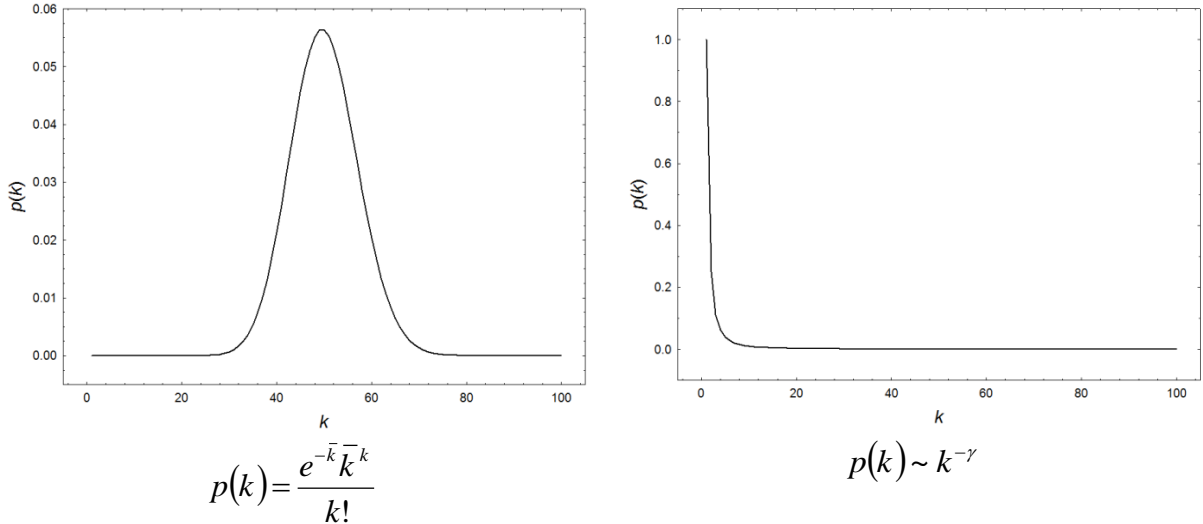

Figure 10.1: Illustration of the Poisson and power-law degree distributions found in complex networks.

In the case of power-law distributions (see Fig. 10.1 left), the probability of finding a node with degree $k$ decays as a negative power of the degree: $p(k) \sim k^{-\gamma}$. This means that the probability of finding a high-degree node is relatively small in comparison with the high probability of finding low-degree nodes. These networks are usually referred to as **'scale-free' networks**. The term scaling describes the existence of a power-law relationship between the probability and the node degree: $p(k) = Ak^{-\gamma}$: scaling the degree by a constant factor $c$, only produces a proportionate scaling of the probability:

$$p(k,c) = A(ck)^{-\gamma} = Ac^{-\gamma} \cdot p(k). \tag{10.1}$$

Power-law relations are usually represented in a logarithmic scale, leading to a straight line, $\ln p(k) = -\gamma \ln k + \ln A$, where $-\gamma$ is the slope and $\ln A$ the intercept of the function. Scaling by a constant factor $c$ means that only the intercept of the straight line changes but the slope is exactly the same as before: $\ln p(k,c) = -\gamma \ln k - \gamma Ac$.

Determining the degree distribution of a network is a complicated task. Among the difficulties we can mention the fact that sometimes the number of data points used to fit the distribution is too small and sometimes the data are very noisy. For instance, in fitting power-law distributions, the tail of the distribution, the part which corresponds to high-degrees, is usually very noisy. There are two main approaches in use for reducing this noise effect in the tail of probability distributions. One is the binning procedure, which consists in building a histogram using bin sizes which increase exponentially with degree. The other approach is to consider the cumulative distribution function (CDF) (Clauset *et al.*, 2010). The cumulative distributions of power-law and Poisson distributions are given below:

$$P(k) = \sum_{k'=k}^{\infty} p(k'), \tag{10.2}$$

$$P(k) = e^{-\bar{k}} \sum_{i=1}^{\lfloor k \rfloor} \frac{(\bar{k})^i}{i!}, \tag{10.3}$$

which represents the probability of choosing at random a node with degree greater than or equal to $k$. In the case of power-law degree distributions, $P(k)$ also shows a power-law decay with degree

$$P(k) \sim \sum_{k'=k}^{\infty} k'^{-\gamma} \sim k^{-(\gamma-1)}, \qquad (10.4)$$

which means that we will also obtain a straight line for the logarithmic plot of $P(k)$ versus $k$ in scale-free networks.

## 10.1 'Scale-free' networks

Among the many possible degree distributions existing for a given network the 'scale-free' one is one of the most ubiquitously found. Consequently, it is important to study a model that is able to produce random networks with such kind of degree distribution. That is, a model in which the probability of finding a node with degree $k$ decreases as a power-law of its degree. The most popular of these models is the one introduced by Barabási and Albert (1999), which is described below.

In the Barabási-Albert (BA) model a network is created by using the following procedure. Start from a small number $m_0$ of nodes. At each step add a new node $u$ to the network and connect it to $m \leq m_0$ of the existing nodes $v \in V$ with probability

$$p_u = \frac{k_v}{\sum_w k_w} . \qquad (10.5)$$

We can assume that we start from a connected random network of the Erdös-Rényi type with $m_0$ nodes, $G_{ER} = (V, E)$. In this case the BA process can be understood as a process in which small inhomogeneities in the degree distribution of the ER network growths in time. Another option is the one developed by Bollobás and Riordan (2004) in which it is first assumed that $d = 1$ and that the $i$ th node is attached to the $j$-th one with probability:

$$p_i = \begin{cases} \dfrac{k_j}{1 + \sum_{j=0}^{i-1} k_j} & \text{if } j < i \\ \dfrac{1}{1 + \sum_{j=0}^{i-1} k_j} & \text{if } j = i \end{cases} . \qquad (10.6)$$

Then, for $d > 1$ the network grows as if $d = 1$ until $nd$ nodes have been created and the size is reduced to $n$ by contracting groups of $d$ consecutive nodes into one. The network is now specified by two parameters and we denote it by $BA(n, d)$. Multiple links and self-loops are created during this process and they can be simply eliminated if we need a simple network.

A characteristic of BA networks is that the probability that a node has degree $k \geq d$ is given by:

$$p(k) = \frac{2d(d-1)}{k(k+1)(k+2)} \sim k^{-3}, \qquad (10.7)$$

which immediately implies that the cumulative degree distribution is given by:

$$P(k) \sim k^{-2}. \qquad (10.8)$$

For fixed values $d \geq 1$, Bollobás (2003) has proved that the expected value for the clustering coefficient $\overline{C}$ is given by

$$\bar{C} \sim \frac{d-1}{8}\frac{\log^2 n}{n}, \tag{10.9}$$

for $n \to \infty$, which is very different from the value $\bar{C} \sim n^{-0.75}$ reported by Barabási and Albert (1999) for $d = 2$.

On the other hand, the average path length has been estimated for the BA networks to be as follows (Bollobás and Riordan, 2004):

$$\bar{l} = \frac{\ln n - \ln(d/2) - 1 - \gamma}{\ln \ln n + \ln(d/2)} + \frac{3}{2}, \tag{10.10}$$

where $\gamma$ is the Euler-Mascheroni constant. This means that for the same number of nodes and average degree, BA networks have smaller average path length than their ER analogues. Other alternative models for obtaining power-law degree distributions with different exponents $\gamma$ can be found in the literature (Dorogovtsev, Mendes, 2003). In closing, using this preferential attachment algorithm we can generate random networks which are different from those obtained by using the ER method in many important aspects including their degree distributions, average clustering and average path length.

## 11 Network motifs

The concept of **network motifs** was introduced by Milo *et al.* (2002) in order to characterize recurring, significant patterns in real-world networks (Milo *et al.*, 2002; 2004). A network motif is a subgraph that appears more frequently in a real network than could be expected if the network were built by a random process. In order to measure the statistical significance of a given subgraph the $Z$-*score* is used, which is defined as follows for a given subgraph $i$:

$$Z_i = \frac{N_i^{real} - \langle N_i^{random} \rangle}{\sigma_i^{random}}, \tag{11.1}$$

where $N_i^{real}$ is the number of times the subgraph $i$ appears in the real network, $\langle N_i^{random} \rangle$ and $\sigma_i^{random}$ are the average and standard deviation of the number of times that $i$ appears in the ensemble of random networks, respectively. Some of the motifs found by Milo *et al.* (2002) in different real-world directed networks are illustrated in the Figure 11.1.

| Motif | Network |
|---|---|
| 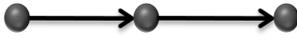 Three-chain | • Food webs |
| 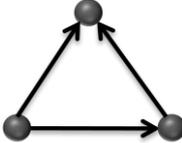 Feedforward loop | • Gene regulation (transcription) <br> • Neurons <br> • Electronic circuits |
| 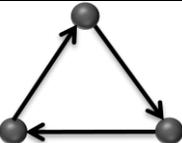 | • Electronic circuits |

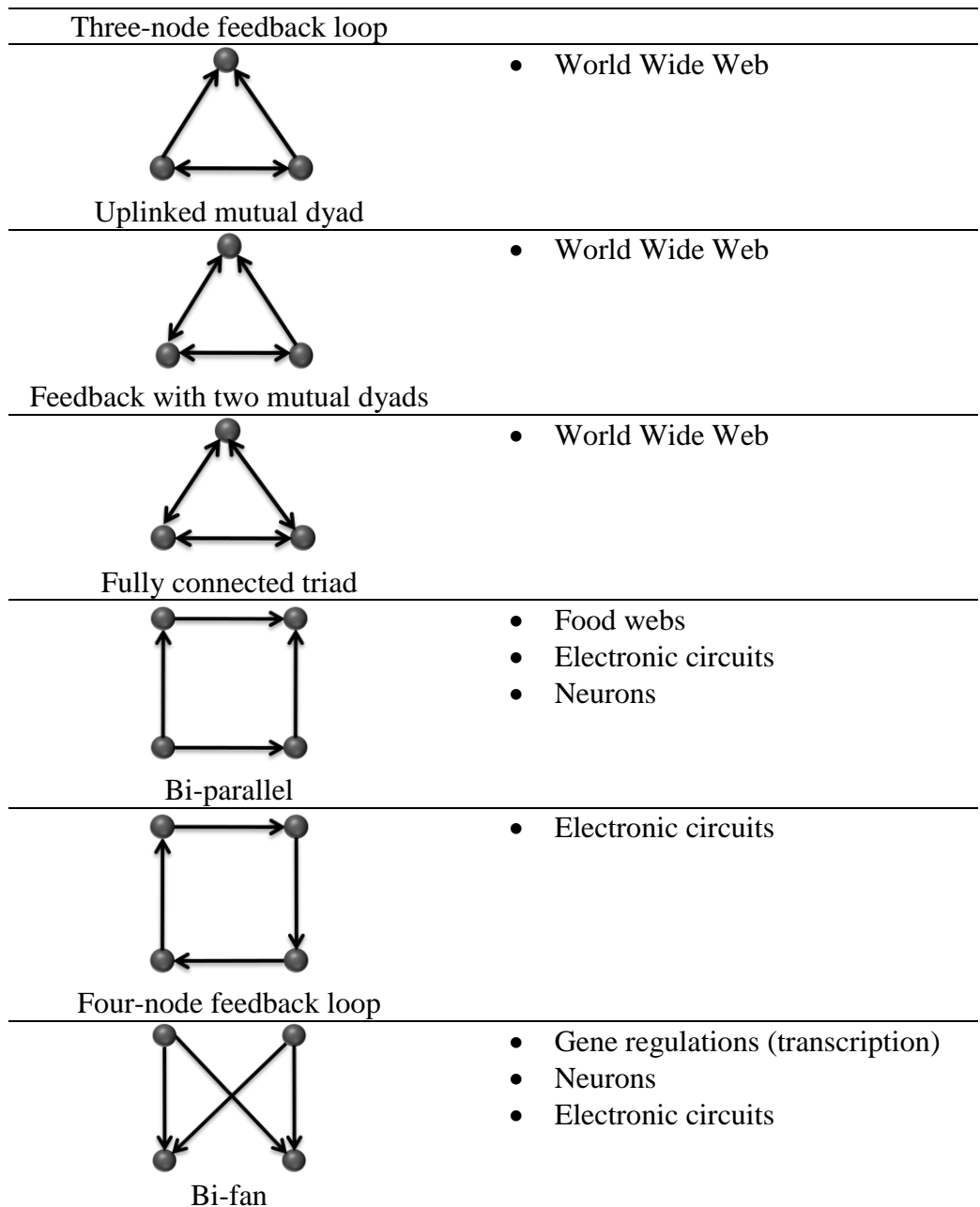

Figure 11.1: Illustration of some of the motifs found in real-world networks.

## 12 Centrality measures

Node centrality in a network is one of the many concepts that have been created in the analysis of social networks and then imported to the study of any kind of networked system. Measures of centrality try to capture the notion of 'importance' of nodes in networks by quantifying the ability of a node to communicate directly with other nodes, or its closeness to many other nodes or the number of pairs of nodes which need a specific node as intermediary in their communications. Here we describe some of the most relevant centrality measures currently in use for studying complex networks.

The **degree** of a node was defined in the first section. It was first considered as a centrality measure for nodes in a network by Freeman (1979) as a way to account for immediate effects taking place in a network. The degree centrality can be written as

$$k_i = \sum_{j=1}^{n} A_{ij}. \tag{12.1}$$

In directed networks, we have two different kinds of degree centrality, namely **in-** and **out-degree** of a node

$$k_i^{in} = \sum_{j=1}^{n} A_{ij}, \tag{12.2}$$

$$k_j^{out} = \sum_{i=1}^{n} A_{ij}. \tag{12.3}$$

Another type of centrality is the **closeness centrality**, which measures how close a node is from the rest of the nodes in the network. The closeness centrality (Freeman, 1979) is expressed mathematically as follows

$$CC(u) = \frac{n-1}{s(u)}, \tag{12.4}$$

where the distance sum $s(u)$ is

$$s(u) = \sum_{v \in V(G)} d(u,v). \tag{12.5}$$

The **betweenness centrality** quantifies the importance of a node is in the communication between other pairs of nodes in the network (Freeman, 1979). It measures the proportion of information that passes through a given node in communications between other pairs of nodes in the network and it is defined as:

$$BC(k) = \sum_{i} \sum_{j} \frac{\rho(i,k,j)}{\rho(i,j)}, \quad i \neq j \neq k \tag{12.6}$$

where $\rho(i,j)$ is the number of shortest paths from node $i$ to node $j$, and $\rho(i,k,j)$ is the number of these shortest paths that pass through node $k$ in the network.

The **Katz centrality index** for a node in a network is defined as (Katz, 1953)

$$K_i = \left\{ \left[ \left( \mathbf{I} - \eta^{-1} \mathbf{A} \right)^{-1} - I \right] \mathbf{1} \right\}_i, \tag{12.7}$$

where $I$ is the identity matrix, $\eta \neq \lambda_1$ is an attenuation factor ($\lambda_1$ is the principal eigenvector of the adjacency matrix) and **1** is column vector of ones. This centrality index can be considered as an extension of the degree in order to consider the influence not only of the nearest neighbors but also of the most distant ones.

The Katz centrality index can be defined for directed networks:

$$K_i^{out} = \left\{ \left[ \left( \mathbf{I} - \eta^{-1} \mathbf{A} \right)^{-1} - I \right] \mathbf{1} \right\}_i, \tag{12.8}$$

$$K_i^{in} = \left\{ \mathbf{1}^T \left[ \left( \mathbf{I} - \eta^{-1} \mathbf{A} \right)^{-1} - \mathbf{I} \right] \right\}_i. \tag{12.9}$$

The $K_i^{in}$ is a measure of the 'prestige' of a node as it accounts for the importance that a node has due to other nodes that point to it.

Another type of centrality that captures the influence not only of nearest neighbors but also of more distant nodes in a network is the **eigenvector centrality**. This index was introduced by Bonacich (1972; 1987) and is the $i$ th entry of the principal eigenvector of the adjacency matrix

$$\varphi_1(i) = \left(\frac{1}{\lambda_1}\mathbf{A}\varphi_1\right)_i. \qquad (12.10)$$

In directed networks there are two types of eigenvector centralities which can be defined by using the principal right and left eigenvectors of the adjacency matrix:

$$\varphi_1^R(i) = \left(\frac{1}{\lambda_1}\mathbf{A}\varphi_1^R\right)_i, \qquad (12.11)$$

$$\varphi_1^L(i) = \left(\frac{1}{\lambda_1}\mathbf{A}^T\varphi_1^L\right)_i. \qquad (12.12)$$

Right eigenvector centrality accounts for the 'importance' of a node by taking into account the 'importance' of nodes to which it points on. That is, it is an extension of the outdegree concept by taking into account not only nearest neighbors. On the other hand, left eigenvector centrality measures the importance of a node by considering those nodes pointing towards the corresponding node and it is an extension of the indegree centrality. This is frequently referred to as 'prestige' in social sciences contexts.

There is an important difficulty when we try to apply right and left eigenvector centralities to networks where there are nodes having outdegree or indegree equal to zero, respectively. In the first case the nodes pointing to a given node do not receive any score for pointing to it. When the indegree is zero, the left eigenvector centrality or prestige of this node is equal to zero as nobody points to it, even though it can be pointing to some important nodes. A solution for this problem is solved by the following centrality measure.

The **PageRank centrality** measure is the tool used by Google in order to rank citations of web pages in the WWW (Langville, Meyer, 2006). Its main idea is that the importance of a web page should be proportional to the importance of other web pages pointing to it. In other words, the PageRank of a page is the sum of the PageRanks of all pages pointing into it. Mathematically, this intuition is captured by the following definition. The PageRank is obtained by the vector

$$\left(\boldsymbol{\pi}^{k+1}\right)^T = \left(\boldsymbol{\pi}^k\right)^T \mathbf{G}, \qquad (12.13)$$

The matrix $\mathbf{G}$ is defined by

$$\mathbf{G} = \alpha\mathbf{S} + \left(\frac{1-\alpha}{n}\right)\mathbf{11}^T, \qquad (12.14)$$

where $0 \le \alpha \le 1$ is a 'teleportation' parameter, which captures the effect in which a web surfer abandons his random approach of bouncing from one page to another and initiates a new search simply by typing a new destination in the browser's URL command line. The matrix $\mathbf{S}$ solve the problem of dead end nodes in ranking web pages, and it is defined as

$$\mathbf{S} = \mathbf{H} + \mathbf{a}\left[(1/n)\mathbf{1}^T\right], \qquad (12.15)$$

where the entries of the dangling vector $\mathbf{a}$ are given by

$$a_i = \begin{cases} 1 & \text{if } k_i^{out} = 0 \\ 0 & \text{otherwise} \end{cases}. \qquad (12.16)$$

Finally, the matrix $\mathbf{H}$ is defined as a modified adjacency matrix for the network:

$$H_{ij} = \begin{cases} 1/k_i^{out} & \text{if threis a link from } i \text{ to } j \\ 0 & \text{otherwise} \end{cases}$$

The matrix $\mathbf{G}$ is row-stochastic which implies that its largest eigenvalue is equal to one and the *principal left-hand eigenvector* of $\mathbf{G}$ is given by

$$\boldsymbol{\pi}^T = \boldsymbol{\pi}^T\mathbf{G}, \qquad (12.17)$$

where $\pi^T \mathbf{1} = 1$ (Langville and Meyer, 2006).

Another type of node centrality which is based on spectral properties of the adjacency matrix of a graph is the **subgraph centrality**. The subgraph centrality counts the number of closed walks starting and ending at a given node, which are mathematically given by the diagonal entries of $\mathbf{A}^k$. In general terms the subgraph centrality is a family of centrality measures defined on the basis of the following mathematical expression:

$$f_i(\mathbf{A}) = \left( \sum_{l=0}^{\infty} c_l \mathbf{A}^l \right)_{ii}, \tag{12.18}$$

where coefficients $c_l$ are selected such that the infinite series converges. One particularly useful weighting scheme is the following, which eventually converges to the exponential of the adjacency matrix (Estrada and Rodríguez-Velazquez, 2005):

$$EE(i) = \left( \sum_{l=0}^{\infty} \frac{\mathbf{A}^l}{l!} \right)_{ii} = \left( e^{\mathbf{A}} \right)_{ii}. \tag{12.19}$$

We can also define subgraph centralities that take into account only contributions from odd or even closed walks in the network:

$$EE_{odd}(i) = (\sinh \mathbf{A})_{ii}, \tag{12.20}$$
$$EE_{even}(i) = (\cosh \mathbf{A})_{ii}. \tag{12.21}$$

A characteristic of the subgraph centrality in directed networks is that it accounts for the participation of a node in directed walks. This means that the subgraph centrality of a node in a directed network is $EE(i) > 1$ only if there is at least one closed walk that starts and returns to this node. In other case $EE(i) = 1$. That is, the subgraph centrality in a directed network measures the returnability of 'information' to a given node.

# 13 Statistical mechanics of networks

Let us consider that every link of a network is weighted by a parameter $\beta$. Evidently, the case $\beta = 1$ corresponds to the simple network. Let $\mathbf{W}$ be the adjacency matrix of this homogeneously weighted network. It is obvious that $\mathbf{W} = \beta \mathbf{A}$ and the spectral moments of the adjacency matrix are $M_r(\mathbf{W}) = \text{Tr}\,\mathbf{W}^r = \beta^r \text{Tr}\mathbf{A}^r = \beta^r M_r$. Let us now count the total number of closed walks in this weighted network. It is straighforward to realize (Estrada and Hatano, 2007) that it is given by

$$Z(G;\beta) = Tr \sum_{r=0}^{\infty} \frac{\beta^r \mathbf{A}^r}{r!} = Tr e^{\beta \mathbf{A}} = \sum_{j=1}^{n} e^{\beta \lambda_j}. \tag{13.1}$$

Let us now consider that the parameter $\beta = (k_B T)^{-1}$ is the inverse temperature of a thermal bath in which the whole network is submerged to. Here the temperature is a physical analogy for the external 'stresses' to which a network is continuously exposed to. For instance, let us consider the network in which nodes represent corporations and the links represent their business relationships. In this case the external stress can represent the economical situation of the World at the moment in which the network is analyzed. In 'normal' economical situations we are in presence of a low level of external stress. In situations of economical crisis the level of external stress is elevated.

We can consider the probability that the network is in a configuration (state) with an energy given by the eigenvalue $\lambda_j$. The configuration or state of the network can be considered here as provided by the corresponding eigenvector of the adjacency matrix associated with $\lambda_j$. This probability is then given by

$$p_j = \frac{e^{\beta\lambda_j}}{\sum_j e^{\beta\lambda_j}} = \frac{e^{\beta\lambda_j}}{Z(G;\beta)}, \qquad (13.2)$$

which identifies the normalization factor as the partition function of the network. This index introduced by Estrada (2000) is known in the graph theory literature as the Estrada index of the graph/network and usually denoted by $EE(G)$.

We can define the **entropy** for the network,

$$S(G;\beta) = -k_B \sum \left[ p_j \left( \beta\lambda_j - \ln Z \right) \right], \qquad (13.3)$$

where we wrote $Z(G;\beta) = Z$ for the sake of economy.

The **total energy** $H(G)$ and Helmholtz free energy $F(G)$ of the network, respectively (Estrada and Hatano, 2007), are given by

$$H(G,\beta) = -\frac{1}{EE} \sum_{j=1}^{n} \left( \lambda_j e^{\beta\lambda_j} \right) = -\frac{1}{EE} \text{Tr}\left( \mathbf{A} e^{\beta\mathbf{A}} \right) = -\sum_{j=1}^{n} \lambda_j p_j, \qquad (13.4)$$

$$F(G,\beta) = -\beta^{-1} \ln EE. \qquad (13.5)$$

Known bounds for the physical parameters defined above, are:
$$0 \leq S(G,\beta) \leq \beta \ln n, \qquad (13.6)$$
$$-\beta(n-1) \leq H(G,\beta) \leq 0, \qquad (13.7)$$
$$-\beta(n-1) \leq F(G,\beta) \leq -\beta \ln n, \qquad (13.8)$$

where the lower bounds are obtained for the complete graph as $n \to \infty$ and the upper bounds are reached for the null graph with $n$ nodes (Estrada and Hatano, 2007).

Next, let us analyse the thermodynamic functions of networks for extreme values of the temperature. At very low temperatures, the total energy and Helmholtz free energy are reduced to the interaction energy of the network (Estrada and Hatano, 2007): $H(G,\beta \to \infty) = F(G,\beta \to \infty) = -\lambda_1$. At very high temperatures, $\beta \to 0$, the entropy of the system is completely determined by the partition function of the network, $S(G;\beta \to 0) = k_B \ln Z$ and $F(G,\beta \to 0) \to -\infty$.

The introduction of these statistical mechanics parameters allows the study of interesting topological and combinatorial properties of networks by using a well-understood physical paradigm. For finding examples of these applications the reader is referred to the specialized literature (Estrada, 2011).

## 13.1 Communicability in networks

The concept of network communicability is a very recent one. However, it has found applications in many different areas of network theory (Estrada et al., 2012). This concept captures the idea of correlation in a physical system and translates it into the context of network theory. We define here that the **communicability** between a pair of nodes in a network depends on all routes that connect these two nodes (Estrada et al., 2012). Among all these routes, the shortest path is the one making the most important contribution as it is the most 'economic' way of connecting two nodes in a network. Thus we can use the weighted sum of all walks of different lengths between a pair of nodes as a measure of their communicability. That is,

$$G_{pq} = \sum_{k=0}^{\infty} \frac{(\mathbf{A}^k)_{pq}}{k!} = (e^{\mathbf{A}})_{pq}, \tag{13.9}$$

where $e^{\mathbf{A}}$ is the matrix exponential function. We can express the communicability function for a pair of nodes in a network by using the eigenvalues and eigenvectors of the adjacency matrix:

$$G_{rs} = \sum_{j=1}^{n} \varphi_j(r)\varphi_j(s)e^{\lambda_j}. \tag{13.10}$$

By using the concept of inverse temperature introduced above we can also express the communicability function in terms of this parameter (Estrada and Hatano, 2008)

$$G_{rs}(\beta) = \sum_{k=0}^{\infty} \frac{(\beta\mathbf{A}^k)_{rs}}{k!} = (e^{\beta\mathbf{A}})_{rs}. \tag{13.11}$$

Intuitively, the communicability between the two nodes connected by a path should tend to zero as the length of the path tends to infinity. In order to show that this is exactly the case we can write the expression for $G_{rs}(\beta)$ for the path $P_n$:

$$G_{rs} = \frac{1}{n+1}\left(\sum_j \cos\frac{j\pi(r-s)}{n+1} - \cos\frac{j\pi(r+s)}{n+1}\right)e^{2\cos\left(\frac{j\pi}{n+1}\right)}, \tag{13.12}$$

where we have used $\beta \equiv 1$ without any loss of generality. Then, it is straightforward to realize by simple substitution in (158) that $G_{rs} \to 0$ for the nodes at the end of a linear path as $n \to \infty$. At the other extreme we find the complete network $K_n$, for which

$$G_{rs} = \frac{e^{n+1}}{n} + e^{-1}\sum_{j=2}^{n} \varphi_j(r)\varphi_j(s) = \frac{e^{n+1}}{n} - \frac{1}{ne} = \frac{1}{ne}(e^n - 1), \tag{13.13}$$

which means that $G_{rs} \to \infty$ as $n \to \infty$. In closing, the communicability measure quantifies very well our intuition that communication decays to zero when only one route exists for connecting two nodes at an infinite distance (the end nodes of a path) and it tends to infinity when there are many possible routes of very short distance (any pair of nodes in a complete graph).

## 14 Communities in networks

The study of **communities** in complex networks is a large area of research with many existing methods and algorithms. The aim of all of them is to identify subsets of nodes in a network the density of whose connections is significantly larger than the density of connections between them and the rest of the nodes. It is impossible to give a complete survey of all the methods for detecting communities in networks in this short section. Thus we are going to describe some of the main characteristics of a group of methods currently used for detecting communities in networks. An excellent review with details on the many methods available can be found in (Fortunato, 2010).

The first group of methods used for detecting communities in networks is that of partitioning methods. Their aim is to obtain a partition of the network into $p$ disjoint sets of nodes such that:

i) $\bigcup_{i=1}^{p} V_i = V$ and $V_i \cap V_j = \phi$ for $i \neq j$,

ii) the number of edges crossing between subsets (cut size or **boundary**) is minimized,

iii) $|V_i| \approx n/p$ for all $i = 1, 2, \ldots, p$, where the vertical bars indicates the cardinality of the set,

When condition (iii) is fulfilled the corresponding partition is called *balanced*. There are several algorithms that have been tested in the literature for the purpose of network partitioning which include local improvement methods and spectral partitioning. The last family of partition methods is based on the adjacency, Laplacian or normalized Laplacian matrices of graphs. In general, their goal is to find a separation between the nodes of the network based on the eigenvectors of these matrices. This separation is carried out *grosso modo* by considering that two nodes $v_1, v_2$ are in the same partition if $\operatorname{sgn} \varphi_2^M(v_1) = \operatorname{sgn} \varphi_2^M(v_2)$ for $M = A, L, \tilde{L}$. Otherwise they are considered to be in two different partitions $\{V_1, V_2\}$. Sophisticated versions of these methods exist and the reader is referred to the specialized literature for details (Fortunato, 2010).

The second group of methods is based on edge centralities. We have defined a number of centrality measures for nodes in a previous section of this Chapter; these can be extended to edges of a network in a straightforward way. In these methods the aim is to identify edges which connect different communities. The best known technique is based on edge betweenness centrality defined for edges in as similar way as for nodes. This method, known as the Girvan-Newman algorithm (Girvan, Newman, 2002), can be summarised in the following steps:

i) Calculate the edge betweenness centrality for all links in the network;
ii) Remove the link with the largest edge betweenness or any of them if more than one exists;
iii) Recalculate the edge betweenness for the remaining links;
iv) Repeat until all links have been removed;
v) Use a dendrogram for analysing the community structure of the network.

Using this dendrogram, a hierarchy of different communities is identified, which can be discriminated by using different quality criteria. The most popular among these quality criteria is the so-called modularity index. In a network consisting of $n_V$ partitions, $V_1, V_2, \ldots, V_{n_C}$, the modularity is the sum over all partitions of the difference between the fraction of links inside each partition and the expected fraction by considering a random network with the same degree for each node (Newman, 2006):

$$Q = \sum_{k=1}^{n_C} \left[ \frac{|E_k|}{m} - \left( \frac{\sum_{j \in V_k} k_j}{2m} \right)^2 \right], \qquad (14.1)$$

where $|E_k|$ is the number of links between nodes in the $k$-th partition of the network. Modularity is interpreted in the following way. If $Q = 0$, the number of intra-cluster links is not bigger than the expected value for a random network. Otherwise, $Q = 1$ means that there is a strong community structure in the network given by the partition analyzed.

The third group of community detection methods is based on **similarity measures** for the nodes in a network. Such similarity measures for the nodes of a network can be based on either rows or columns of the adjacency matrix of the network. For instance, we can consider as a measure of similarity between two nodes the angle between the corresponding rows or columns of these two nodes in the adjacency matrix of the graph. This angle is defined as

$$\sigma_{ij} = \cos \vartheta_{ij} = \frac{\mathbf{x}^T \mathbf{y}}{\|\mathbf{x}\| \cdot \|\mathbf{y}\|}, \tag{14.2}$$

which can be seen to equal

$$\sigma_{ij} = \frac{\eta_{ij}}{\sqrt{k_i k_j}}, \tag{14.3}$$

where $\eta_{ij}$ is the number of common neighbours of nodes $i$ and $j$.

Other similarity measures between rows or columns of the adjacency matrices are the Pearson correlation coefficient, different types of norms and distances (Manhattan, Euclidean, infinite), etc. Once a similarity measure has been chosen, any of the variety of similarity-based methods for detecting communities in a network can be used.

# 15 Dynamical processes on networks

There are many dynamical processes that can be defined on graphs and networks. The reader should be aware that this is a vast area of multidisciplinary research with a huge number of publications in different fields.

## 15.1 Consensus

We will start here with a simple model for analyzing consensus among the nodes in a network. We consider a graph $G = (V, E)$ whose nodes represent agents in a complex system and the edges represent interactions between such agents. In such multi-agent system, "**consensus**" means an agreement regarding a certain quantity of interest. An example is a collection of autonomous vehicles engaged in cooperative teamwork in civilian and military applications. Such coordinated activity allows them to perform missions with greater efficacy than if they perform solo missions (Olfati-Saber et al., 2007).

Let $n = |V|$ be the number of agents forming a network, the collective dynamics of the group of agents is represented by the following equations for the continuous-time case:

$$\dot{\boldsymbol{\varphi}} = -\mathbf{L}\boldsymbol{\varphi}, \quad \boldsymbol{\varphi}(0) = \boldsymbol{\varphi}_0, \tag{15.1}$$

where $\boldsymbol{\varphi}_0$ is the original distribution which may represent opinions, positions in space or other quantities with respect to which the agents should reach a consensus. The reader surely already recognized that Eq. (15.1) is identical to the heat equation,

$$\frac{\partial u}{\partial t} = h \Delta u, \tag{15.2}$$

where $h$ is a positive constant and $\nabla^2 = -\mathbf{L}$ is the Laplace operator. In general this equation is used to model the diffusion of 'information' in a physical system, where by information we can understand heat, a chemical substance or opinions in a social network.

A consensus is reached if, for all $\varphi_i(0)$ and all $i, j = 1, \ldots, n$, $|\varphi_i(t) - \varphi_j(t)| \to 0$ as $t \to 0$. The discrete-time version of the model has the form

$$\varphi_i(t+1) = \varphi_i(t) + \varepsilon \sum_{j \sim i} \mathbf{A}_{ij} \left[ \varphi_j(t) - \varphi_i(t) \right], \ \varphi(0) = \varphi_0 \quad (15.3)$$

where $\varphi_i(t)$ is the value of a quantitative measure on node $i$, $\varepsilon > 0$ is the step-size, and $j \sim i$ indicates that node $j$ is connected to node $i$. It has been proved that the consensus is asymptotically reached in a connected graph for all initial states if $0 < \varepsilon < 1/\delta_{max}$, where $\delta_{max}$ is the maximum degree of the graph. The discrete-time collective dynamics of the network can be written in matrix form as (Olfati-Saber et al., 2007) as

$$\varphi(t+1) = \mathbf{P}\varphi(t), \ \varphi(0) = \varphi_0, \quad (15.4)$$

where $\mathbf{P} = \mathbf{I} - \varepsilon \mathbf{L}$, and $\mathbf{I}$ is the $n \times n$ identity matrix. The matrix $\mathbf{P}$ is the Perron matrix of the network with parameter $0 < \varepsilon < 1/\delta_{max}$. For any connected undirected graph the matrix $\mathbf{P}$ is an irreducible, doubly stochastic matrix with all eigenvalues $\mu_j$ in the interval $[-1, 1]$ and a trivial eigenvalue of 1. The reader can find the previously mentioned concepts in any book on elementary linear algebra. The relation between the Laplacian and Perron eigenvalues is given by: $\mu_j = 1 - \varepsilon \lambda_j$.

In Figure 15.1 we illustrate the consensus process in a real-world social network having 34 nodes and 78 edges.

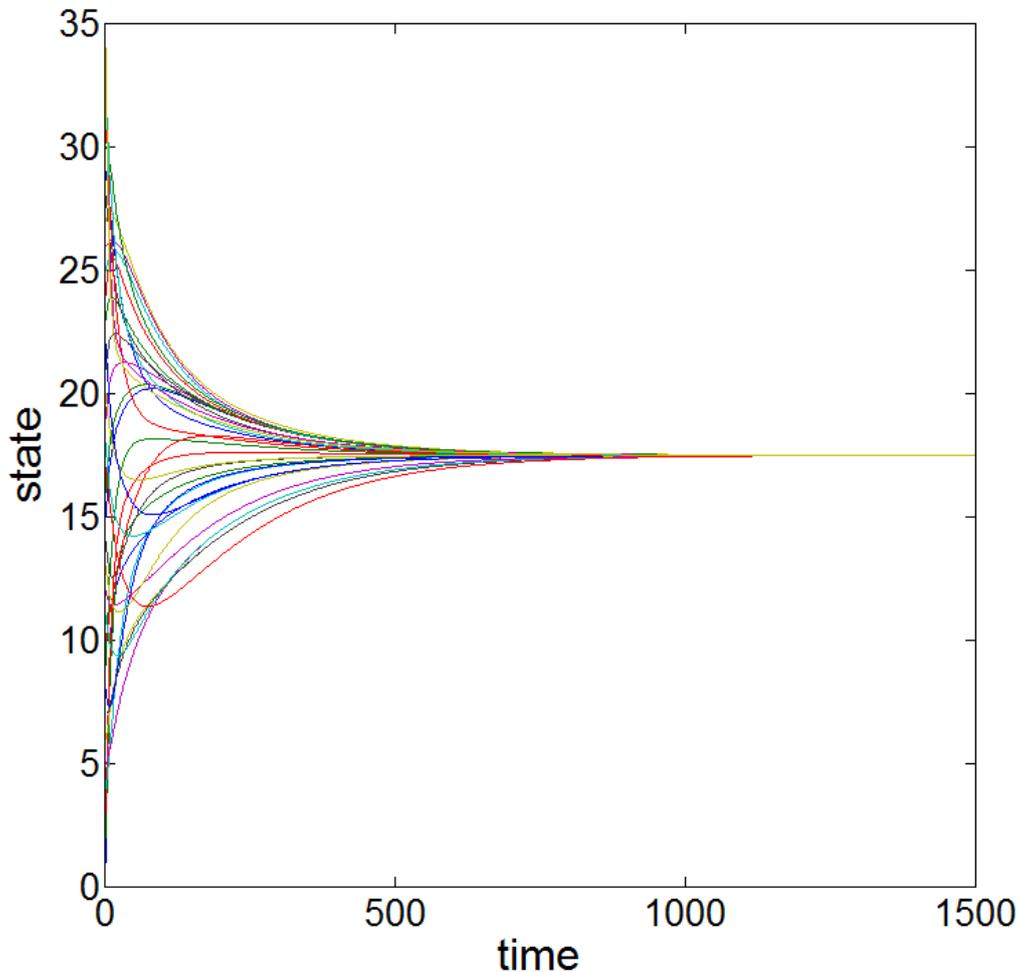

Figure 15.1: Time evolution of consensus dynamics in a real-world social network with random initial states for the nodes.

## 15.2 Synchronization in networks

A problem which is closely related to that of consensus in networks is one of **synchronization** (Arenas et al., 2006; Cheng et al., 2009). The phenomenon of synchronization appears in many natural systems consisting of a collection of oscillators coupled to each other. These systems include animal and social behavior, neurons, cardiac pacemaker cells, among others. We can start by considering a network $G = (V, E)$ with $|V| = n$ nodes representing coupled identical oscillators. Each node is an $N$-dimensional dynamical system which is described by the following equation

$$\dot{x}_i = f(x_i) + c \sum_{j=1}^{n} L_{ij} H(t) x_j, \quad i = 1, \ldots, n, \tag{15.5}$$

where $x_i = (x_{i1}, x_{i2}, \ldots, x_{iN}) \in \mathbb{R}^N$ is the state vector of the node $i$, $f(\cdot) : \mathbb{R}^N \to \mathbb{R}^N$ is a smooth vector valued function which defines the dynamics, $c$ is a constant representing the coupling strength, $H(\cdot) : \mathbb{R}^N \to \mathbb{R}^N$ is a fixed output function also known as outer coupling matrix, $t$ is the time and $L_{ij}$ are the elements of the Laplacian matrix of the network (sometimes the negative of the $H(x_i) \approx H(s) + \xi_i H'(s)$ Laplacian matrix is taken here). The network is said to achieve synchronization if

$$x_1(t) = x_2(t) = \cdots = x_n(t) \to s(t), \quad \text{as} \quad t \to \infty. \tag{15.6}$$

Let us now consider a small perturbation $\xi_i$ such that $x_i = s + \xi_i$ ($\xi_i \ll s$) and let us analyze the stability of the synchronized manifold $x_1 = x_2 = \ldots = x_n$. First, we expand the terms in (15.5) as

$$f(x_i) \approx f(s) + \xi_i f'(s), \tag{15.7}$$

$$H(x_i) \approx H(s) + \xi_i H'(s), \tag{15.8}$$

where the primes refers to the derivatives respect to $s$. Thus, the evolution of the perturbations is determined by the following equation:

$$\dot{\xi}_i = f'(s) \xi_i + c \sum_j \left[ L_{ij} H'(s) \right] \xi_j. \tag{15.9}$$

It is known that the system of equations for the perturbations can be decoupled by using the set of eigenvectors of the Laplacian matrix, which are an appropriate set of linear combinations of the perturbations. Let $\phi_j$ be an eigenvector of the Laplacian matrix of the network associated with the eigenvalue $\mu_j$. Recall that the Laplacian is positive semi-definite, i.e., $0 = \mu_1 \leq \mu_2 \leq \ldots \leq \mu_n \equiv \mu_{max}$. Then

$$\dot{\phi}_i = \left[ f'(s) + c \mu_i H'(s) \right] \phi_i. \tag{15.10}$$

Let us now assume that at short times the variations of $s$ are small enough to allow us to solve these decoupled equations, with the solutions being

$$\dot{\phi}_i(t) = \phi_i^0 \exp\left\{\left[f'(s) + c\mu_i H'(s)\right]t\right\}, \qquad (15.11)$$

where $\phi_i^0$ is the initially impossed perturbation.

We now consider the term in the exponential of (15.11), $\Lambda_i = f'(s) + c\mu_i H'(s)$. If $f'(s) > c\mu_i H'(s)$ the perturbations will increase exponentially, while if $f'(s) < c\mu_i H'(s)$ they will decrease exponentially. So, the behavior of the perturbations in time is controlled by the magnitude of $\mu_i$. Then, the stability of the synchronized state is determined by the master stability function:

$$\Lambda(\alpha) \equiv \max_s \left[f'(s) + \alpha H'(s)\right], \qquad (15.12)$$

which corresponds to a large number of functions $f$ and $H$ is represented in the Figure 15.2.

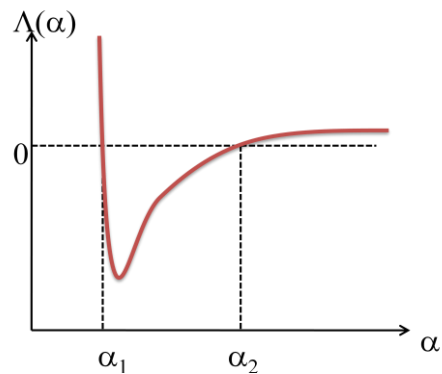

Figure 15.2: Schematic representation of the typical behavior of the master stability function.

As can be seen the necessary condition for stability of the synchronous state is that $c\mu_i$ is between $\alpha_1$ and $\alpha_2$, which is the region where $\Lambda(\alpha) < 0$. Then, the condition for synchronization is (Barahona, Pecora, 2002):

$$Q := \frac{\mu_N}{\mu_2} < \frac{\alpha_2}{\alpha_1}. \qquad (15.13)$$

That is, synchronizability of a network is favored by a small eigenratio $Q$ which indeed depends only on the topology of the network. There are many studies on the synchronizability of networks using different types of oscillators and the reader is referred to the specialized literature for the details (See Further reading, Barrat, Barthelemy, Vespignani, 2008).

## 15.3 Epidemics on networks

Another area in which the dynamical processes on networks play a fundamental role is the study of the spread of epidemics. These models are extensions of the classical models used in epidemiology which consider the influence of the topology of a network on the propagation of an epidemic (Keeling, Eames, 2005). The simplest model assumes that an individual who is susceptible (S) to an infection could become infected (I). In a second model the infected individual can also recover (R) from infection. The first model is known as a SI model, while the second is known as a SIR model. In a third model, known as SIS, an individual can be reinfected, so that infections do not confer immunity on an infected individual. Finally, a model known as SIRS allows for recovery and reinfection as an attempt to model the temporal immunity conferred by certain infections. Here we briefly consider only the SIR and SIS models on networks.

In the **SIR** (*susceptible-infected-recovered*) model there are three compartments as sketched in the Figure 15.3:

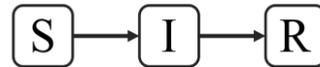

Figure 15.3: Diagrammatic representation of a SIR model.

That is, in a network $G = (V, E)$ a group of nodes $S \subseteq V$ are considered susceptible and they can be infected by directd contact with infected individuals. Let $s_i$, $x_i$ and $r_i$ be the probabilities that the node $i$ is susceptible, infected or has recovered. The evolution of these probabilities in time is governed by the following equations that define the **SIR model**:

$$\dot{s}_i = -\beta s_i \sum_j A_{ij} x_j ,  \quad (15.14)$$

$$\dot{x}_i = \beta s_i \sum_j A_{ij} x_j - \gamma x_i ,  \quad (15.15)$$

$$\dot{r}_i = \gamma x_i ,  \quad (15.16)$$

where $\beta$ is the spreading rate of the pathogen, $A_{ij}$ is an entry of the adjacency matrix of the network and $\gamma$ is the probability that a node recovers or dies, i.e., the recovery rate.

In the **SIS** (susceptible-infected-susceptible) model the general flow chart of the infection can be represented as in Figure 15.4:

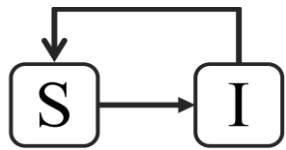

Figure 15.4: Diagrammatic representation of a SI model.

The equations governing the evolution of the probabilities of susceptible and infected individuals are given below:

$$\dot{s}_i = -\beta s_i \sum_j A_{ij} x_j + \gamma x_i ,  \quad (15.17)$$

$$\dot{x}_i = \beta s_i \sum_j A_{ij} x_j - \gamma x_i .  \quad (15.18)$$

The analysis of epidemics in networks is of tremendous importance in modern life. Today, there is a large movility of people across cities, countries and the entire world and an epidemic can propagate through the social networks at very high rates. The reader can find a few examples in the specialized literature (Keeling and Eames, 2005).

# Glossary

**adjacency matrix** – of a simple graph: a binary symmetric matrix whose row and columns represent the vertices of the graph, where the $i, j$ entry is one if the corresponding vertices $i$ and $j$ are connected.
**betweenness centrality** – a centrality measure for a node which characterizes how central a node is in passing information from other nodes.
**bipartite graph** – a graph with two sets of vertices, the nodes of each set being connected only to nodes of the other set.
**bridge** – an edge whose deletion increases the number of connected components of the graph.
**centrality measure** – an index for a node or edge of a graph/network that characterizes its topological or structural importance.
**closeness centrality** – a centrality measure for a node which characterizes how close the node is respect to the rest in terms of the shortest-path distance.
**clustering coefficient** – the ratio of the number of trigles incident to a node in a graph to the maximum possible number of such triangles.
**communicability** – a measure of how well-communicated a pair of nodes is by considering all possible routes of communication in a graph/network.
**complete graph** – a graph in which every pair of vertices are connected to each other.
**connected graph** – a graph in which there is a path connecting every pair of nodes.
**cycle** – a path in which the initial and end vertices coincide.
**cycle graph** – a graph in which every node has degree two.
**degree** – a centrality measure for a node which counts the number of edges incident to a node.
**degree distribution** – the statistical distribution of the degrees of the nodes of a graph.
**edge contraction** – a graph operation in which an edge of the graph is removed and the two end nodes are merged together.
**edge deletion** – a graph operation in which an edge of the graph is removed leaving the end nodes in the graph.
**Erdös-Rényi graph** – a random graph formed from a given set of nodes and a probability of create edges among them.
**forest** – a graph formed by several components all of which are trees.
**girth** – the size of the minimum cycle in a graph.
**graph** – a pair formed by a set of vertices or nodes and a set of edges.
**graph diameter** – the length of the largest shortes-path distance in a graph.
**graph diameter** – the maximum shortest path distance in a graph.
**graph invariant** – a characterization of a graph which does not depends on the labelling of vertices or edges.
**graph nullity** – the multiplicity of the zero eigenvalue of the adjacency matrix, i.e., the number of times eigenvalue zero occurs in the spectrum of the adjacency matrix.

**hydrocarbon** – a molecule formed only by carbon and hydrogen.
**incidence matrix** – of a graph: a matrix whose rows correspond to vertices and whose columns correspond to edges of the graph and the $i, j$ entry is one or zero if the $i$ th vertex is incident with the $j$ th edge or not, respectively.
**Laplacian matrix** – a square symmetric matrx with diagonal entries equal to the degree of the corresponding vertex and off-diagonal entries equal to -1 or zero depending if the corresponding vertices are connected or not, respectively.
**loop** – an edge which is doubly incident to the same node.
**matching of a graph** – the number of mutually non-adjacent edges in the graph.
**mean displacement** - of an atom (vertex): refers to the oscillations of an atoms from its equilibrium position due to thermal fluctuations.
**molecular Hamiltonian** – is the operator representing the energy of the electrons and atomic nuclei in a molecule.
**network community** – a subset of nodes in a graph/network which are better connected among them than with the rest of the nodes.
**network motif** – a subgraph in a graph which is overrepresented in relation to a random graph of the same size.
**path** - a sequence of different consecutive vertices and edges in a graph.
**path graph** – a tree in which all nodes have degree two except two nodes which has degree one.
**regular graph** – a graph in which every node has the same degree.
**resistance distance** – a distance between any pair of vertices of the graph, determined by the Kirchhoff rules for electrical sets.
**scale-free network** – a network/graph with a power-law degree distribution.
**shortest path** – between two nodes: a path having the least number of edges among all paths connecting two vertices.
**simple graph** – a graph without multiple edges, self-loops and weights.
**spanning forest** – a subgraph of graph which containg all the nodes of the graph and is a forest.
**spanning tree** – a subgraph of a graph which contain all the nodes of the graph and is also a tree.
**star graph** – a tree consisting of a node with degree n-1 and n-1 nodes with degree one.
**tree** – a graph that does not have any cycle.
**vertex degree** – number of vertices adjacent to a given vertex.
**walk** – a sequence of (not necessarily) different consecutive vertices and edges in a graph.

## List of works cited

Arenas, A., Diaz-Guilera, A., Pérez-Vicente, C. J. (2006). *Physica D* **224**, 27-34.
Bapat, R. B., Gutman, I., Xiao, W. (2003). *Z. Naturforsch.* **58a**, 494 – 498.
Barabási, A.-L. Albert, R. (1999). *Science* **286**, 509-512.
Barahona, M., Pecora, L. M. (2002). *Phys. Rev. Lett.* **89**, 054101.
Barrat, A., Weigt, M. (2000). *Eur. Phys. J. B.* **13**, 547-560.
Beaudin, L., Ellis-Monaghan, J., Pangborn, G., Shrock, R. (2010). *Discr. Math.* **310**, 2037-2053.
Berkolaiko, G., Kuchment, P. (2013). *Introduction to Quantum Graphs* (Vol. 186). Amer Mathematical Society.

Biggs, N. L., Lloyd, E. K., Wilson, L. (1976). *Graph Theory 1736-1936*. Clarendon Press, Oxford.

Bogner, C. (2010). *Nucl. Phys. B-Proc. Supp.* **205**, 116-121.

Bogner, C., Weinzierl, (2010). *Int. J. Mod. Phys. A* **25**, 2585.

Bollobás, B., (1998). *Modern Graph Theory*. Springer-Verlag, New York.

Bollobás, B., (2003). In Bornholdt, S., Schuster, H. G. (Eds.), *Handbook of Graph and Networks: From the genome to the internet*, Wiley-VCH, Weinheim, pp. 1-32.

Bollobás, B., Riordan, O. (2004). *Combinatorica* **24**, 5-34.

Bonacich, P. (1972). *J. Math. Sociol*. **2**, 113–120.

Bonacich, P., (1987). *Am. J. Soc*. **92**, 1170-1182.

Borovićanin, B., Gutman, I. (2009). In *Applications of Graph Spectra,* D. Cvetković and I. Gutman (Eds.), Math. Inst. SANU, 2009, pp. 107-122.

Canadell, E., Doublet, M.-L., Iung, C. (2012). *Orbital Approach to the Electronic Structure of Solids*. Oxford University Press, Oxford.

Chen, G., Wang, X., Li, X., Lü, J. (2009). In *Recent Advances in Nonlinear Dynamics and Synchronization*, K. Kyamakya (Ed.), Springer-Verlag, Berlin, pp. 3-16.

Cheng, B., Liu, B. (2007). *Electron. J. Linear Algebra* 16 (2007), 60-67.

Clauset, A., Rohilla Shalizi, C., Newman, M. E. J. (2010). *SIAM Rev*. **51**, 661-703.

Dodgson, C. L. (1866). *Proc. Roy. Soc. London* **15**, 150-155.

Doyle, P., Snell, J. (1984). *Random Walks and Electric Networks*. Carus Math. Monogr. 22, Washington, D.C.

Dorogovtsev, S. N., Mendes, J. F. F. (2003). *Evolution of Networks: From Biological Nets to the Internet and WWW*. Oxford University Press, Oxford.

Ellis-Monaghan, J. A., Merino, C. (2011). In *Structural Analysis of Complex Networks*, M. Dehmer (ed.), pp. 219-255.

Erdös, P., Rényi, A. (1959). *Publ.Math. Debrecen* **5**, 290-297.

Essam, J. W. (1971). *Discr. Math.* **1**, 83-112.

Estrada, E. (2000). *Chem.Phys. Lett*. **319**, 713-718.

Estrada, E., Hatano, N. (2007). *Chem. Phys. Let*. **439**, 247-251.

Estrada, E., Hatano, N. (2008). *Phys. Rev. E* **77**, 036111.

Estrada, E., Hatano, N., Benzi, M., (2012). *Phys. Rep*. **514**, 89-119.

Estrada, E., Rodríguez-Velázquez, J. A. (2005). *Phys. Rev. E* **71**, 056103.

Euler L. (1736). *Comm. Acad. Scient. Imp. Petrop.*, **8**, 128-140.

Fortunato, S. (2010). *Phys. Rep*. **486**, 75-174.

Foss, S., Korshunov, D., Zachary, S. (2011). *An Introduction to Heavy-Tailed and Subexponential Distributions*. Springer, Berlin.

Freeman, L. C. (1979). *Social Networks* **1**, 215–239.

Ghosh, A., Boyd, S., Saberi, A. (2008). *SIAM Rev*. **50**, 37-66.

Girvan, M., Newman, E. J. (2002). *Proc. Natl. Acad. Sci. USA* **99**, 7821-7826.

Gutman, I. (2005). *J. Serbian Chem. Soc.* 70, 441-456.

Gutman, I., Xiao, O. (2004). *Bull. Acad. Serb. Sci. Arts*. **29**, 15-23.

Harary, F. (1969). *Graph Theory*. Addison-Wesley, Reading, MA.

Harary, F. (Ed.) (1968). *Graph Theory and Theoretical Physics*. Academic Press Inc.,US.

Janson, S. (2005). *J. Combin. Prob. Comput.* **14**, 815-828.

Katz, L. (1953). *Psychometrica* **18**, 39-43.

Keeling, M. J., Eames, K. T. (2005). *J. Roy. Soc. Interface* **2**, 295-307.

Klein, D. J., Randić, M. (1993). *J. Math. Chem.* **12**, 81-95.

Kutzelnigg, W. (2006). *J. Comp. Chem.* 28, 25-34.

Langville, A. N., Meyer, C. D. (2006). *Google's PageRank and Beyond. The Science of Search Engine Rankings*. Princeton University Press, Princeton.

Lieb, E. H. (1989). *Phys. Rev. Lett.* **62**, 1201 (erratum (1989) **62,** 1927).

Milgram, S. (1967). *Psychol. Today* **2**, 60-67.

Milo, R., Itzkovitz, S., Kashtan, N., Levitt, R., Shen-Orr, Shai., Ayzenshtat, I., Sheffer, M., Alon, U. (2004). *Science* **303**, 1538-1542.

Milo, R., Shen-Orr, S., Itzkovitz, S. Kashtan, N. Chklovskii, D. Alon, U. (2002). *Science* **298**, 824-827.

Morita, Y., Suzuki, S., Sato, K., Takui, T. (2011). *Nature Chem.* **3**, 197-204.

Newman, M. E. J. (2006). *Proc. Natl. Acad. Sci. USA* **103**, 8577-8582.

Newman, M. E. J., Strogatz, S. H., Watts, D. J. (2001). *Phys. Rev. E* **64**, 026118.

Olfati-Saber, R., Fax, J. A., Murray, R. M. (2007) *Proc. IEEE* **95**, 215-233

Ottino, J. M. (2003). *AIChE J.*, **49**, 292-299.

Powell, B. J. (2009). An introduction to effective low-energy Hamiltonians in condensed matter physics and chemistry. *arXiv preprint arXiv:0906.1640*.

Standish, R. K. (2008). Concept and definition of complexity. In *Intelligent Complex Adaptive Systems*, Yang, A. and Shan, Y. (eds), (IGI Global: Hershey, PA) pp105-124, arXiv:0805.0685.

Sylvester, J. J. (1877-1878). *Nature* **17**, 284-285.

Tasaki, H. (1999). *J. Phys.: Cond. Mat.*, **10**, 4353.

Trinajstić, N. (1992). *Chemical Graph Theory*. CRC Press, Boca Raton, FL.

Watts, D. J., Strogatz, S. H. (1998). *Nature* **393**, 440-442.

Weinzierl, S. (2010). Introduction to Feynman integrals. *arXiv preprint arXiv:1005.1855*.

Welsh, D. (1999). *Random Struct. Alg.*, **15**, 210–228.

Welsh, D. J. A., Merino, C. (2000). *J. Math. Phys.* **41**.

Wigner, E. P. (1955). *Ann Math.* **62**, 548-564.

Xiao, W., Gutman, I., (2003). *Theor. Chem. Acc*. **110**, 284-289.

**Further reading**

Barrat, A., Barthélemy, M., Vespignani, A. (2008). *Dynamical Processes on Complex Networks*. Cambridge University Press, Cambridge.

Bollobás, B. (1998). *Modern Graph Theory*. Springer, Berlin.

Caldarelli, G. (2007). *Scale-Free Networks. Complex Webs in Nature and Technology*. Oxford University Press, Oxford.

Cvetković, D., Rowlinson, P., Simić, S. (2010). *An Introduction to the Theory of Graph Spectra*. Cambridge University Press, Cambridge.

Estrada, E. (2011). *The Structure of Complex Networks. Theory and Applications*. Oxford University Press, Oxford.

Nakanishi, N. (1971). *Graph Theory and Feynman Integrals*. Gordon and Breach.